\documentclass[acmlarge]{acmart}
\makeatletter
\newcommand{\confshort}{\acmConference@shortname}
\newcommand{\conffull}{\acmConference@name}
\newcommand{\confdate}{\acmConference@date}
\newcommand{\confloc}{\acmConference@venue}
\AtBeginDocument{
  \fancypagestyle{firstpagestyle}{
    \fancyhead{}%
    \fancyfoot[C]{}%
  }
  \fancyhf{}
  \fancyhead[LO]{\@headfootfont\shorttitle}%
  \fancyhead[RE]{\@headfootfont\@shortauthors}%
  \fancyhead[LE]{\@headfootfont\footnotesize \confshort, \confdate, \confloc}%
  \fancyhead[RO]{\@headfootfont\footnotesize \confshort, \confdate, \confloc}%
  \fancyfoot[C]{}%
}
\makeatother
\acmBooktitle{\conffull\@ (\confshort), \confdate, \confloc}

\usepackage{enumitem}
\usepackage{stfloats}
\usepackage{graphicx}
\usepackage{tcolorbox}
\usepackage{balance}
\usepackage[utf8]{inputenc}
\usepackage{fontawesome5} 
\usepackage{newunicodechar}
\newunicodechar{‪}{}
\newunicodechar{‬}{}
\usepackage{arydshln}
\usepackage{fancyhdr}

\AtBeginDocument{%
  }

\setcopyright{acmlicensed}
\copyrightyear{2026}
\acmYear{2026}
\acmDOI{10.1145/3805689.3812393}
\acmConference[FAccT '26]{The 2026 ACM Conference on Fairness, Accountability, and Transparency}{June 25--28, 2026}{Montreal, Canada}

\acmISBN{979-8-4007-2596-8/2026/06}

\begin{document}

\title[How Hyper-Datafication Impacts the Sustainability Costs in Frontier AI]{How Hyper-Datafication Impacts the Sustainability Costs in \\ Frontier AI}

\author{Sophia N. Wilson}
\orcid{0000-0002-4960-8223}
\affiliation{%
  \institution{University of Copenhagen}
  \city{Copenhagen}
  \country{Denmark}
}
\email{sophia.wilson@di.ku.dk}

\author{Sebastian Mair}
\orcid{0000-0003-2949-8781}
\affiliation{%
  \institution{Linköping University}
  \city{Linköping}
  \country{Sweden}
}
\email{sebastian.mair@liu.se}

\author{Mophat Okinyi}
\orcid{0009-0000-5932-3232}
\affiliation{%
  \institution{Techworker Community Africa}
  \city{Nairobi}
  \country{Kenya}
}
\email{okinyi.mophat@gmail.com}

\author{Erik B. Dam}
\orcid{0000-0002-8888-2524}
\affiliation{%
  \institution{University of Copenhagen}
  \city{Copenhagen}
  \country{Denmark}
}
\email{erikdam@di.ku.dk}

\author{Janin Koch}
\orcid{0000-0001-9207-9550}
\affiliation{%
  \institution{Univ. Lille, Inria, CNRS, Centrale Lille
  }
  \city{Lille}
  \country{France}
}
\email{janin.koch@inria.fr}

\author{Raghavendra Selvan}
\orcid{0000-0003-4302-0207}
\affiliation{%
  \institution{University of Copenhagen}
  \city{Copenhagen}
  \country{Denmark}
}
\email{raghav@di.ku.dk}

\renewcommand{\shortauthors}{Wilson et al.}

\begin{abstract}
Large-scale data has fuelled the success of frontier artificial intelligence (AI) models over the past decade. This expansion has relied on sustained efforts by large technology corporations to aggregate and curate internet-scale datasets. In this work, we examine the environmental, social, and economic costs of large-scale data in AI through a sustainability lens. We argue that the field is shifting from {\em building models from data} to actively {\em creating data for building models}. We characterise this transition as {\em hyper-datafication}, which marks a critical juncture for the future of frontier AI and its societal impacts. To quantify and contextualise data-related costs, we analyse approximately 550,000 datasets from the Hugging Face Hub, focusing on dataset growth, storage-related energy consumption and carbon footprint, and societal representation using language data.
We complement this analysis with qualitative responses from data workers in Kenya to examine the labour involved, including direct employment by big tech corporations and exposure to graphic content. We further draw on external data sources to substantiate our findings by illustrating the global disparity in data centre infrastructure. 
Our analyses reveal that hyper-datafication drives substantial and growing environmental costs while systematically redistributing labour risks and representational harms toward the Global South.
Thus, we propose Data PROOFS recommendations spanning provenance, resource awareness, ownership, openness, frugality, and standards 
to mitigate these costs. Our work aims to make visible the often-overlooked costs of data that underpin frontier AI and to stimulate broader debate within the research community and beyond.
\looseness=-1

\end{abstract}



\maketitle

\section{Introduction} 

The recent advancements in artificial intelligence (AI) have enabled automation that was widely considered decades away. 
AI systems now deliver breakthroughs across scientific discovery, language understanding, and generative modelling, addressing long-standing research challenges while opening new applications~\cite{silver2017mastering,brown2020language,brandt2020unexpectedly,jumper2021highly,lauritzen2022artificial,lam2023learning}. 
This progress is driven primarily by deep learning methods~\cite{lecun2015deep,schmidhuber2015deep}, including large language models (LLMs), multimodal generative models, and reasoning models~\cite{radford2018improving,brown2020language,team2023gemini,grattafiori2024llama,guo2025deepseek}. 

While algorithmic and hardware improvements have played an important role, {\em scale} has acted as the dominant catalyst~\cite{kaplan2020scaling,hoffmann2022training,sevilla2022compute}. 
Frontier AI development now relies on unprecedented levels of resources: models with hundreds of billions of parameters, hyper-scale data centre infrastructure, electricity consumption comparable to that of entire towns, and financial investments that exceed the gross domestic product of several countries~\cite{raghav2025sustainable-ai}. 

\begin{figure}[t]
    \centering
    \includegraphics[width=\linewidth]{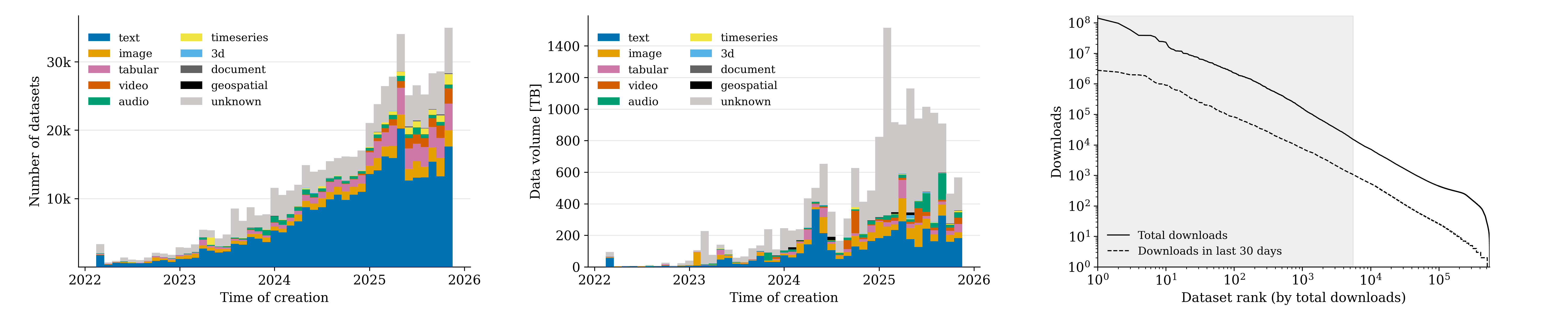}
    \caption{Growth of datasets and data volume over time and download concentration on the Hugging Face Hub. Left: Monthly counts of newly created datasets (100\% of datasets included). Centre: Monthly data volume added (91\% of datasets included). The bars are coloured by modality. For multimodal datasets, each modality contributes proportionally to the colouring. 
    Right: Download concentration. The plot shows total downloads (solid line) and downloads in November 2025 -- which is 30 days preceding metadata extraction -- (dashed line) against dataset rank sorted by total downloads on a logarithmic scale. 
    The shaded region highlights the top 1\% of datasets (5,543 repositories). 
    }
    \Description{Dataset growth and usage on the Hugging Face Hub. The figure illustrates rapid expansion in the number and volume of datasets, where text datasets contribute the largest share, alongside a highly skewed distribution of downloads, where a small share of datasets receives by far the most usage. }
    \label{fig:datasets_and_data_over_time_bars}
\end{figure}

These investments would be futile without access to another critical resource: \emph{data}. Contemporary frontier AI models train on datasets containing hundreds of trillions of data points often scraped at scale from the internet. 
For example, one of the largest publicly available tokenised datasets {\tt DCLM-Pool}, contains more than 240 trillion tokens~\cite{li2024datacomp}.
Figure~\ref{fig:datasets_and_data_over_time_bars} illustrates the rapid growth of datasets and data volume using the datasets available on the Hugging Face Hub\footnote{\url{https://huggingface.co/datasets}}, a public platform for hosting AI models and datasets (see Section~\ref{sec:methods}). 

Growing concern about the sustainability of AI has prompted an expanding body of work~\cite{van2021sustainable,wu2022sustainable,raghav2025sustainable-ai}. 
However, existing discussions typically focus on a single dimension of sustainability and largely centre on model training and deployment. The sustainability of data itself remains comparatively under-examined, despite the fact that collecting and using internet-scale datasets for frontier AI is highly resource-intensive and can generate substantial environmental, social, and economic costs. As a result, current sustainability discussions overlook data as a central driver of AI's broader impacts. 

In this work, we take a comprehensive look at the sustainability of data for AI. 
Building on existing critiques of the sustainability of AI~\cite{van2021sustainable,wu2022sustainable,wright2023efficiency}, we shift the focus from models to data and explicitly consider the environmental, social, and economic costs together. 
This joint perspective reflects that these costs are inherently interdependent and must be understood within the broader context of holistic sustainability. Optimising one dimension in isolation risks shifting burdens elsewhere.
This perspective becomes increasingly urgent as frontier AI research pursues ever more ambitious goals, including artificial general intelligence (AGI), intensifying the demand for data at scale. 

At the same time, however, access to large-scale data has enabled substantial societal benefits. In climate modelling, data-driven systems can now predict climate-related disasters such as floods and wildfires with high accuracy~\cite{jailani2025hybrid, papakis2025multimodal}, improving early warning systems and disaster preparedness. In health care, large-scale data has enabled breakthroughs in drug-discovery, including protein-folding~\cite{jumper2021highly}, supports earlier disease detection~\citep{litjens2016deep,kooi2017large,lauritzen2022artificial} and more personalised treatment strategies~\citep{fung2020achieving,gu2022personalized}. Such applications demonstrate the transformative potential of large-scale data to directly save and improve human lives. 

While these benefits are real and significant, we argue that the field is undergoing a structural shift in how data is produced and used.
Increasingly, large-scale data is not only leveraged to address concrete societal challenges, but is created primarily to sustain model scaling. One manifestation of this shift is the growing reliance on synthetic data in frontier AI systems.
For example, the Phi-4 model was pretrained on approximately 10~trillion tokens, of which around 40\% were synthetically generated and curated using other AI models~\cite{abdin2024phi}. 
More fundamentally, AI development is driving the active creation of new data sources designed explicitly for model training.  
A prominent example is the Ego4D dataset, which consists of 3,670~hours of egocentric video collected from 923~participants across 74~locations, recorded specifically to support AI systems~\cite{grauman2022ego4d}. 
Such datasets would not exist outside the demands of model development. 

We use the term {\em hyper-datafication} to characterise these trends, which we define formally below. While the term has appeared sporadically in prior work, it has been used narrowly to describe the large-scale expansion and aggregation of existing data sources~\citep{Gruyter_handbook_2024}. We extend this usage by explicitly incorporating two additional, increasingly central dimensions: the algorithmic generation of synthetic data and the creation of data whose primary purpose is to serve as training material for AI models rather than to support direct human activities.

\begin{tcolorbox}[colback=blue!10, colframe=blue!10]
{\bf Hyper-datafication} refers to the industrialised production and accumulation of data for AI model development across three coupled processes: (i) the large-scale data collection and recombination of existing data sources, 
(ii) the use of AI systems to generate synthetic data, and 
(iii) the creation of purpose-built data whose primary function is to serve as training input for AI systems rather than direct human use.
\looseness=-1
\end{tcolorbox}

\noindent Assessing the sustainability costs of hyper-datafication requires examining how large-scale data production, storage, and downstream use translate into environmental, social, and economic costs. Despite the central role of data in frontier AI systems, these dimensions remain only partially documented (see Section~\ref{sec:related}). To analyse these costs, we draw on multiple complementary sources: a large-scale metadata study of datasets stored on the Hugging Face Hub; empirical evidence from data workers in Kenya; and external reports and public data to contextualise global data-centre investments and infrastructure. The main contributions of this work are:

\begin{itemize}
    \item {\bf A large-scale empirical analysis of more than 550,000 datasets stored on the Hugging Face Hub}, documenting growth trends, storage- and carbon-related energy consumption for both provider and user sides, and patterns of linguistic representation relative to global speaker populations and web presence.
    \item {\bf Qualitative evidence from 134 data workers in Kenya}, revealing conditions of low pay, lack of compensation for exposure to graphic content, direct employment relationships with large technology companies, and gender-disaggregated disparities.  
    \item {\bf An analysis of the global investments and geographical distribution of data-centres}, drawing on external data sources and public reports to document disparities in data centre infrastructure investment that concentrate economic power among a small number of countries. 
    \item {\bf The Data PROOFS recommendations}, a framework outlining actions required to increase data sustainability and reduce the negative impacts of hyper-datafication. 
\end{itemize}

The following section briefly discusses existing research on the sustainability impacts of AI systems and the importance of data in this context (Section~\ref{sec:related}). 
We then describe our methods for gathering and analysing information on the sustainability implications of AI-related data (Section~\ref{sec:methods}), before presenting our findings in light of the growing trend of hyper-datafication (Section~\ref{sec:costs}). We conclude with a set of recommendations under the Data PROOFS framework (Section~\ref{sec:discussion}). 
\looseness=-1

\section{Background and Related Work}
\label{sec:related}

\subsection{Sustainability of AI}
\label{sec:sust-of-ai}

Sustainability is broadly understood to have three pillars: environmental, social, and economic~\cite{purvis2019three}. 
Together, these reflect the need to balance long-term economic viability, ecological limits, and societal well-being when assessing systemic impacts. This holistic approach is essential for addressing the global challenges when pursuing resource-intensive technologies like AI on a planet with a rapidly changing climate~\cite{kaack2022aligning}. 

{\bf Energy Consumption and Carbon Footprint of AI.} The emerging critique on the sustainability of AI has been primarily focused on its environmental impact; particularly, the growing carbon footprint due to the energy consumption of AI~\cite{strubell2019energy,anthony2020carbontracker,henderson2020towards}. The International Energy Agency (IEA) projects that data centre electricity consumption worldwide will more than double by 2030 (to between 945 and 1300~TWh as depicted in Figure~\ref{fig:datacenter_electricity}), mainly driven by AI workloads~\citep{hecweb2025, unep2024}. This is evident in recent hyper-scale data centre projects. For instance, Microsoft's USD 106 billion data centre with 3.33 GW capacity in Wisconsin, the United States (US), will consume electricity equivalent to that of 3.3 million households in Wisconsin~\cite{EpochAIDataCenters2025} (see Appendix~\ref{app:wisconsin} for details). Globally, energy production continues to be one of the largest sources of greenhouse gas (GHG) emissions~\cite{bruckner2014energy,iea2025global}. 
The growing energy needs of data centres for AI result in a corresponding carbon footprint that has also risen in recent years~\cite{strubell2019energy,anthony2020carbontracker,freitag2021real,luccioni2023estimating}.

{\bf Broader Environmental Impact of AI.} 
Data centres use vast volumes of fresh water for cooling~\cite{li2025making}
threatening freshwater resources in vulnerable regions~\citep{eesi_datacenters_water}. The manufacture of hardware, use of rare earths, electronic waste (e-waste) generation, and land-use change associated with AI data infrastructure contribute major embodied carbon emissions and ecological damage~\citep{kaack2022aligning,wright2023efficiency}. The amount of e-waste has grown from 34 million tons in 2010 to 64 million tons in 2022, and is projected to grow to 82 million tons by 2030. During the same time, the fraction of e-waste recycled decreased from 24\% to less than 5\%~\citep{gem2024}. 

{\bf Social Sustainability of AI.} The social cost of AI has been studied using the lens of fairness and bias mitigation~\cite{barocas2023fairness}, differential privacy~\cite{dwork2006calibrating,abadi2016deep}, and security using robustness to adversarial attacks~\cite{cohen2019certified}. Recent studies have begun to examine the labour required to build frontier AI models~\cite{cant2024feeding,rudolph2025hidden}, a critical factor that is often overlooked in popular debates regarding the social impact of artificial intelligence~\cite{pasquinelli2023eye}. Finally, the democratisation of AI is negatively impacted due to the large-scale resource consumption and concentration of these expensive resources with a few actors~\cite{van2021sustainable,ahmed2020democratization,bakhtiarifard2025climate}.

\subsection{Costs of Data}
The majority of the literature discussed in Section~\ref{sec:sust-of-ai} focuses on model selection, model training, and model deployment costs. The underlying data costs are subsumed into model development. However, this does not adequately capture the actual resource costs. For example, the model card for the 405 billion parameter open-source LLM, Llama-3.1, reports a carbon footprint of 8,930 tonnes of carbon dioxide equivalent (tCO$_2$eq) due to the energy consumption of 21.5 GWh~\cite{llama3modelcard,raghav2025sustainable-ai}. This only includes the training cost and does not include any data-related energy consumption. Curating the dataset for training Llama-3.1 with about 15 trillion tokens, generating synthetic data of 25 million tokens, preprocessing and storing them also incur additional environmental costs. When aggregated across the full scale of data used for AI, these costs can be substantial. 

{\bf Environmental.} While there are no comprehensive studies that estimate the environmental cost of AI datasets, there are specialised studies that focus on domain-specific data. For example, \citet{souter2025comparing} study the impact of data preprocessing in medical image analysis, benchmarking the effects of different preprocessing stages. In Wang et al., authors estimate the direct and embodied emissions due to global land cover mapping projects which is a data-intensive task due to the high-resolution satellite images~\cite{wang2025carbon}. They estimated the emissions between 2014 and 2024 to be about 3.89 million tCO$_2$eq, and project it to increase to 184 million tCO$_2$eq by 2050. Outside of AI, carbon emissions across the data lifecycle have been explored by \citet{mersy2024toward}; they propose {\em carbon provenance} to annotate data with carbon emission meta-data, facilitating better carbon accounting of data.

{\bf Economic.} Research on the economic consequences of AI data highlights the uneven distribution of the burdens and the benefits across regions and actors. Studies show that AI development reproduces structural inequalities between the Global South and the Global North~\citep{mohamed2020decolonial, trabelsi2024impact}. The extreme market concentration reinforces this inequality: a small group of dominant platforms holds disproportionate control over behavioural and operational data, creating data monopolies and high barriers to entry for smaller actors~\citep{zuboff2023age, schweitzer2019competition}.
By analysing the economic data value chain, \citet{jia2025a} argue that monetary value systematically accrues to aggregators and model developers, i.e., the dominant platforms, while data generators are largely excluded.

{\bf Social.}
Parallel research examines the social costs borne by data workers who produce, clean, and label data. 
Studies of annotation centres show that labour is governed through continuous measurement of speed, volume, and accuracy with strict productivity expectations and constant pressure to meet daily targets~\citep{chandhiramowuli2024making}. Interviews with annotators in India and Madagascar further document the pervasive surveillance and escalating productivity demands~\citep{le2023problem,wang2022whose}. Research on micro-task platforms reports similar patterns, including long and irregular working hours, unpaid work, and significant income volatility~\citep{hornuf2022hourly,tubaro2022hidden}. Beyond these conditions, the mental health impacts of data work are increasingly recognised. Studies of content moderation report high levels of stress, secondary trauma, and PTSD-like symptoms among workers exposed to disturbing material~\citep{spence2023psychological, spence2025content, steiger2021psychological}. While much of this research has focused on social-media moderation, the documented harms also extend to content moderation within AI data pipelines.

Frontier AI development is relying on large-scale datasets with trillions of data points~\cite{sevilla2022compute,llama3modelcard,epoch2022trendsintrainingdatasetsizes}. There are even speculations that we will ``run out of data'', prompting the use of synthetic data for AI development~\cite{villalobos2024position}. Considering the costs of such hyper-datafication is crucial to address the sustainability of frontier AI comprehensively. 
\looseness=-1

\section{Methodology}
\label{sec:methods}
To address the costs of hyper-datafication holistically, we estimate the energy use of large-scale data storage (environmental), analyse labour conditions using questionnaires administered to data workers and patterns of language representation in AI datasets (social), and draw on public data to examine economic costs associated with AI data (economical).
This type of empirical analysis necessarily requires grounding in concrete cases and we therefore draw on a set of complementary data sources. First, the Hugging Face Hub serves as our primary lens on dataset growth trends. As one of the most widely used public repositories in AI development, it combines broad community adoption with detailed, consistent metadata, and has been used in meta-analyses in prior studies~\cite{di2024automated,castano2024analyzing}. Second, we use data workers in Kenya as a case study for data labour costs, as Kenya has been identified as a leading hub for AI-related gig work~\cite{wef2025trade}, frequently described as the ``Silicon Savannah''~\cite{graham2013imagining}.
Finally, we draw on the IEA~\cite{iea2025} and Data Center Map \cite{datacentermap} to contextualise the economic scale and distribution of data centre infrastructure. Together, these sources serve as strong indicators of broader trends, although each carries limitations that we discuss in Section~\ref{sec:discussion}.
\looseness=-1

Drawing on these sources, we aim to identify: (i)~the environmental impacts of data storage under hyper-datafication; (ii)~the actors and labour conditions that sustain hyper-datafication; (iii)~the extent to which hyper-datafication represents society using language as a proxy; and (iv)~the economic burdens associated with hyper-datafication.

\subsection{Metadata of Hugging Face Datasets} 
To analyse the trends in dataset growth, quantify storage-related energy costs, and examine language representation, we analyse metadata for datasets hosted on the Hugging Face Hub. 

{\bf Scope.} We retrieved metadata on 1 December 2025, when the Hugging Face Hub hosted 570,802 datasets. 
Metadata extraction failed for 2.89\% of repositories, yielding a final sample of 554,300 datasets. 
We collected repository-level metadata (identifiers, timestamps, downloads, and Hub-side storage), dataset-level metadata (dataset size), and contextual attributes (region, modality, task, and language). Hugging Face has options to query download statistics in two granularities: all time and last 30 days. As we retrieved our data on 1 December 2025, the statistics for {\em last 30 days} correspond to November 2025\footnote{Source code used to extract metadata from datasets hosted on the Hugging Face Hub is available at: \url{https://github.com/saintslab/costs-of-hyperdatafication}.}. 

{\bf Data collection.} We collected individual repository-level metadata via the Hugging Face Hub Representational State Transfer (REST) Application Programming Interface (API)\footnote{Using the endpoint \url{https://huggingface.co/api/datasets/{id}} (one request per dataset ID) and retrieving attributes via the API's \texttt{expand} fields.}, a standard HyperText Transfer Protocol (HTTP) interface that returns structured metadata for a given dataset by its ID. Dataset size statistics were obtained separately from the Hugging Face Hub datasets-server API\footnote{See \url{https://github.com/huggingface/dataset-viewer} for documentation of the datasets-server infrastructure.}, which computes dataset statistics and aggregates results across configurations and splits. Contextual attributes were extracted from the Hugging Face Hub Python API\footnote{Accessing the public dataset catalogue at \url{https://huggingface.co/datasets}.}, a Python client library that wraps the Hub's REST interface and enables programmatic iteration over the full public dataset catalogue.   
These attributes are primarily self-declared and therefore exhibit lower coverage.
Appendix~\ref{app:metadata_coverage} summarises all metadata attributes and coverage. 
\looseness=-1

{\bf Data analysis.} We distinguish two storage measures: dataset size (the size of the raw data files in bytes) and Hub-side storage (the total space occupied on Hugging Face's servers, including additional metadata and auxiliary files beyond the raw data). The former serves as a proxy for local storage footprint, and the latter for platform storage. Language tags were mapped to ISO-639 codes, an international standard for language classification. For the language analysis, dataset volumes were compared against global speaker distributions and web presence. As a proxy for the latter, we used page counts from Common Crawl, a publicly available web archive widely used for training language models. 
Extended descriptive statistics are provided in Appendix~\ref{app:metadata_modalities}.

\subsection{Survey of Data Workers}
To assess social and labour costs, we conducted an online survey targeting data workers in Kenya, administered via a questionnaire. 
\looseness=-1

{\bf Questionnaire design and administration.} 
The questionnaire was designed to address gaps in literature by examining critical aspects of data work, formulated as research questions covering demographics, working conditions, exposure to graphic content, and employment relationships with large technology companies.
It was developed in accordance with established principles of 
questionnaire design~\cite{roopa2012questionnaire} and questions were kept concise and written in plain English to ensure accessibility. 
The questionnaire was developed by members of the research team and independently reviewed by others prior to deployment to ensure clarity and alignment with the study's objectives. 
The final questionnaire comprised ten questions and is provided in Appendix~\ref{app:questionnaire}. 
It was administered via an online form and distributed through a messaging application to data workers who had voluntarily joined the channel organised by a local data workers' collective. 
Data collection took place in December~2025. 

{\bf Participants.} The final sample includes 134 respondents located in Kenya. 
The sample comprises 57~females and 77~males, predominantly aged 20-40 years, with most reporting at least four years of experience. 
Respondents were informed about the study's purpose and how their data would be used. They were not compensated, as institutional policy precludes participant payment. The survey was designed to take around five minutes to complete and the participants were free to participate and stop the survey at any time. 

{\bf Data analysis.} We computed descriptive statistics and pairwise associations across working hours, experience, salary, exposure to graphic content, and employment type. 
Some questions allowed free-text responses, which, in a few cases, were used to harmonise categories. 
Free-text responses that could not be mapped consistently to the categories were excluded. 
Detailed preprocessing decisions and complete raw counts are reported in Appendix~\ref{app:questionnaire_raw_counts}.

\subsection{External Data Sources}
We extract global annual investment in data centre infrastructure from the IEA~\citep{iea2025}. 
To characterise the geographic concentration and expansion of data centres, we use records from the Data Center Map~\citep{datacentermap}. We assessed electricity demand associated with data centres using historical and projected estimates from the IEA~\citep{iea2024energydemandai} and focus on electricity consumption as a primary indicator of environmental impact associated with data centre expansion. We source the carbon intensity for electricity generation from Our World in Data~\cite{owid_carbon_intensity_2025}.

\section{Costs of Hyper-Datafication}
\label{sec:costs}

The costs associated with large-scale data accumulation are multifaceted, and the different dimensions overlap and interact in practice. We separate these categories into environmental, social, and economic for analytical clarity. Moreover, each dimension involves nuances beyond the scope of this study. We therefore focus on aspects most relevant for understanding the structural consequences of sustained data growth.

\subsection{Environmental Costs}
\label{sec:env}

Hyper-datafication creates a persistent obligation to store data. As datasets grow in number and size, storage requirements accumulate and become environmentally significant. Unlike model training and inference, storage remains largely invisible in environmental sustainability discussions, yet it underpins the continued operation of data-intensive AI systems. The Hugging Face Hub provides a concrete case for examining how these storage demands scale over time. 

Figure~\ref{fig:datasets_and_data_over_time_bars} (Left and Centre) shows a sharp increase in both the number and volume of newly created datasets on the Hugging Face Hub since March 2022. Over this period, the monthly dataset upload rate has increased by more than an order of magnitude. 
At the same time, the volume of newly added data has increased by nearly three orders of magnitude, from tens of terabytes per month to peak levels exceeding one petabyte per month.  
While dataset creation has accelerated rapidly, usage remains highly concentrated. 
Figure~\ref{fig:datasets_and_data_over_time_bars} (Right) shows that the top 1\% of datasets (5,543 repositories) account for 87.3\% of total downloads and 81.7\% of downloads in November 2025.
\looseness=-1

\begin{figure}[t]
    \centering
    \includegraphics[width=.9\linewidth]{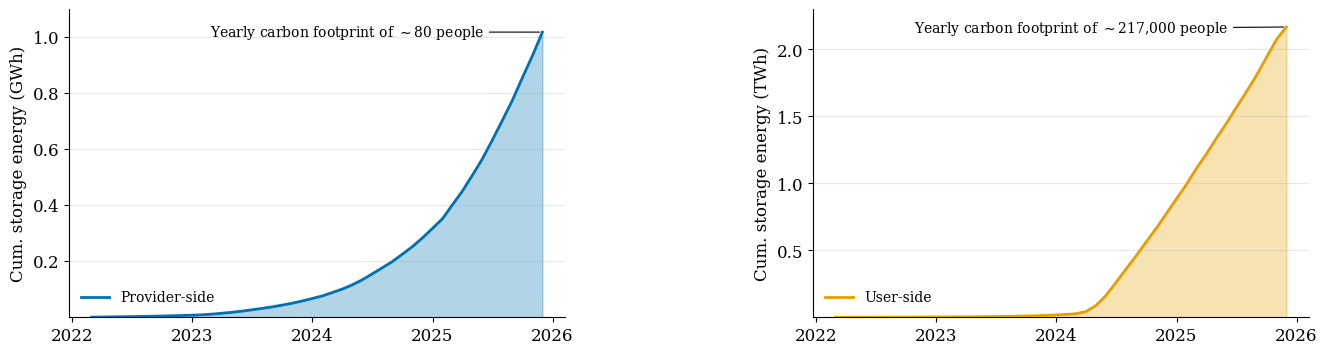}
    \caption{Left: Estimated provider-side storage energy (GWh). Right: Estimated user-side storage energy (TWh), assuming that 10\% of downloads result in three months of local storage. To provide scale, cumulative energy is expressed relative to the global average annual per-capita carbon footprint of 4.73 tCO$_2$eq~\citep{owid_co2_per_capita_2023}. For the provider-side, we use the US grid intensity of 384 gCO$_2$eq/kWh~\citep{owid_carbon_intensity_2025} reflecting that nearly all datasets are hosted in the US, while for the user-side we use the global average of 473 gCO$_2$eq/kWh~\citep{owid_carbon_intensity_2025} due to the lack of geographic information on downloads. Provider-side storage thus corresponds to the annual carbon footprint of approximately 80 people and user-side storage to that of roughly 217,000 people. 
    \looseness=-1}
    \label{fig:storage_energy}
    \Description{Estimated cumulative energy use associated with data storage on the Hugging Face Hub. The figure shows that both provider-side and user-side storage energy increase over time, with user-side storage contributing the dominant share under the stated assumptions.}
\end{figure}

Figure~\ref{fig:storage_energy} reports estimated provider-side (Left) and user-side storage (Right) energy consumption associated with Hugging Face datasets. These estimates cover 91\% of datasets for the provider-side estimate and 75\% for the user-side estimate and is based on assumptions and calculations provided in Appendix~\ref{app:storage-emissions}. 
Provider-side storage energy consumption remains modest in absolute terms but increases steadily as datasets accumulate, having reached approximately 1 GWh.  
User-side costs are approximately three orders of magnitude larger ($>$2 TWh), driven by the repeated number of downloads across a large user base. Here, we have assumed that 10\% of downloads result in three months of persistent storage per user. These choices are conservative and our estimate should therefore be interpreted as a lower bound.

\begin{figure*}[t]
    \centering
    \includegraphics[width=\linewidth]{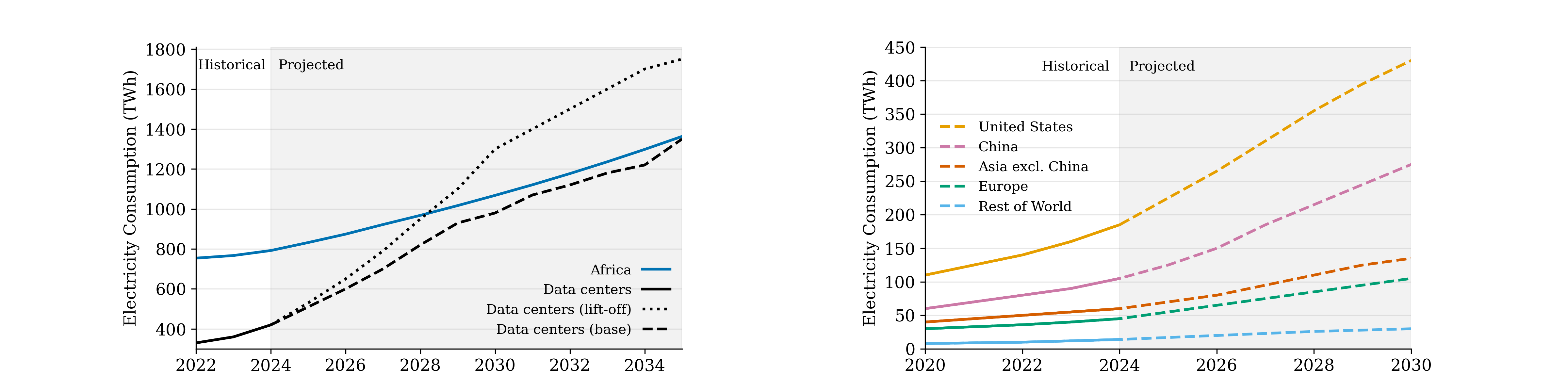}
    \caption{Left: Historical (2022–2024) and projected (2024–2034) electricity use for all data centres worldwide under two scenarios: a base case reflecting current regulatory conditions and industry projections, and a lift-off case assuming stronger AI adoption enabled by faster data centre deployment as modelled by the IEA~\citep{iea2024energydemandai}. Shown alongside the total electricity consumption in Africa assuming a 5\% annual grow~\citep{iea2025electricity}. Right: Regional distribution of data centre electricity demand in the base case (2020–2030), showing contributions from the US, China, Asia excluding China, Europe, and the rest of the world~\citep{iea2024energydemandai}. 
    \looseness=-1
    }
    \label{fig:datacenter_electricity}
    \Description{Historical and projected electricity use by data centres. The figure shows that electricity demand from data centres increases over time under both scenarios. In the lift-off scenario, projected data centre electricity use exceeds Africa’s total electricity consumption after year 2028, while the base scenario remains about the same. Regional projections indicate that growth is steepest in the United States and China.}
\end{figure*}

Since November~2023, Hugging Face has enabled users to specify the geographic storage region for datasets\footnote{\url{https://huggingface.co/docs/hub/storage-regions}}. 
The available regions are the US and the European Union (EU).
Our results show an intense concentration of storage in the US. Of the datasets with region information (99.8\%), 552,713 are hosted in the US, compared with only 316 in the EU. This is not surprising, as the majority of Hugging Face's infrastructure is in the US. 

Carbon intensity varies across electricity grids due to the differences in the energy mix. Storing all datasets hosted on the Hugging Face Hub in the EU rather than the US would reduce the associated carbon footprint by approximately 38\%, due to the lower average carbon intensity in the EU (237 gCO$_2$eq/kWh) compared with the US (384 gCO$_2$eq/kWh) in 2024~\cite{owid_carbon_intensity_2025}. Depending on the location of data centre infrastructure, storage-related carbon footprints can differ by up to a factor of 30 across regions\footnote{This reflects contrasts between high- and low-carbon electricity grids, such as Kosovo (959 gCO$_2$eq/kWh) and Norway (31 gCO$_2$eq/kWh) in 2024~\cite{owid_carbon_intensity_2025}.}. 
\looseness=-1

The user-side carbon footprint of storing local copies of datasets from the Hugging Face Hub is found to be comparable to the annual carbon footprint of about 217,000 people.

Estimating electricity consumption and carbon emissions for {\em all AI datasets} is infeasible beyond focused analyses such as the one presented using the Hugging Face Hub. This has to be performed at a coarser scale using data centre-level analysis. Figure~\ref{fig:datacenter_electricity} (Right) shows the growth projection of data centres, which is highly uneven across geographic regions. The US dominates both historical and projected electricity use, followed by China. Electricity demand in Asia, excluding China, Europe, and the rest of the world, remains substantially lower. Nevertheless, hyper-datafication and increasing pressure to adopt frontier AI drove a 27\% increase in global data centre electricity demand between 2022 and 2024~\cite{iea2024energydemandai}. Projections indicate this consumption will soon rival the total electricity usage of the entire African continent, as shown in Figure~\ref{fig:datacenter_electricity} (Left).
\looseness=-1

In addition to the carbon footprint due to the operational electricity consumption of a data centre, there are additional emissions related to data transmission. Operating networking infrastructure, such as wireless network equipment, optical fibre, and other switching equipment, contributes additional emissions~\cite{malmodin2014life}. However, when considering transmission costs, the embodied emissions from the manufacturing and construction of underwater optical fibre networks are the primary factors. It is reported that the embodied emissions of optical fibre are about 2.3 kgCO$_2$eq/km~\cite{sullivan2025optical} which are amortised over every bit of data that is transferred.

\subsection{Social Costs}
\label{sec:social}

{\bf Data Worker Conditions in Kenya.}  Kenya has become one of the regional hubs for data work, supported by (relatively) strong mobile infrastructure and a young, technologically capable population~\cite{wef2025trade}. 
Current estimates suggest that around 1.9 million Kenyans participate in digital labour, including roughly 1.2 million gig workers. This accounts for approximately 3.3\% of the population, however, the sector contributes more than 9\% of the national GDP~\cite{wef2025trade}. 

We analyse labour-related social costs due to hyper-datafication based on the responses from 134 data workers in Kenya. 
Respondents report engaging in a wide range of tasks, including data labelling (120/134), content moderation (67/134), data cleaning (52/134), and verification of user preferences (33/134), with many performing multiple task types concurrently. 
A majority (79/134) report that they currently work or have previously worked directly for a major technology company, including OpenAI (51/134), Meta (35/134), Google (24/134), Microsoft (13/134), and Amazon (10/134). 
\looseness=-1

\begin{figure}[b]
    \centering
    \includegraphics[width=\linewidth]{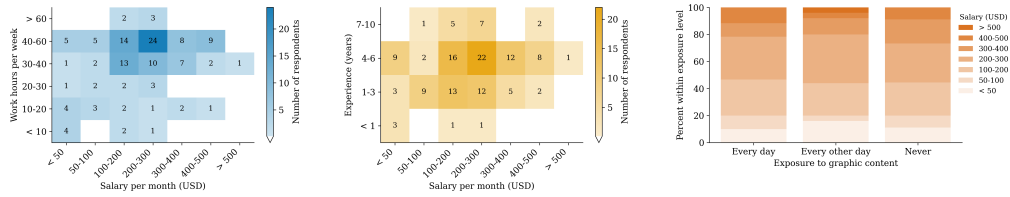}
    \caption{Left: Distribution of respondents across salary bands by weekly working hours. Centre: Distribution of respondents across salary bands by years of experience. Right: Monthly salary distribution by exposure level, indicating no clear relationship between exposure level and salary.}
    \label{fig:workhours_experience_salary}
    \Description{Heatmaps and bar charts showing how salary varies with working hours, experience, and exposure to graphic content among data workers. Most respondents earning higher salaries work longer hours and have more years of experience. By contrast, salary distributions remain similar across daily, occasional, and no exposure to graphic content, with no systematic increase in pay at higher exposure levels.}
\end{figure}

Figure~\ref{fig:workhours_experience_salary} shows the relationship between working hours, experience, exposure, and salary. 
Roughly half of respondents (65/134) report working between 40 and 60 hours per week. 
The most common experience level is 4-6 years (70/134), while the most common salary range is USD 200-300 per month (42/134).
For comparison, the Kenyan national average is USD 540 per month~\citep{CEIC2024KenyaEarnings}.

Exposure to graphic content emerges as a central social cost. 
A majority of respondents (89/134) report some level of exposure, with 60 reporting daily exposure and 25 reporting exposure every other day. 
This exposure is especially common among workers engaged in content moderation, which is increasingly rebranded as ``Trust and Safety''. Contrary to common claims, our data suggests that higher exposure does not correspond to higher pay. Figure~\ref{fig:workhours_experience_salary} (Right) shows no consistent association between exposure level and salary. Conditioning on working hours and experience yields the same result. 

Among respondents who report having worked directly for a large technology company at some point, half (40/79) report daily exposure to graphic content and a further significant share reports exposure every other day (17/79). Among those who do not report working for large technology companies, exposure rates are roughly one-third (20/55) report daily exposure, and a smaller share report exposure every other day (8/55). The questionnaire does not distinguish between current and prior employment, and these patterns should therefore not be interpreted as evidence that direct employment with large technology companies causes higher exposure. 

\begin{figure}[t]
\begin{minipage}{0.79\textwidth}
    \centering
    \includegraphics[width=0.4\linewidth]{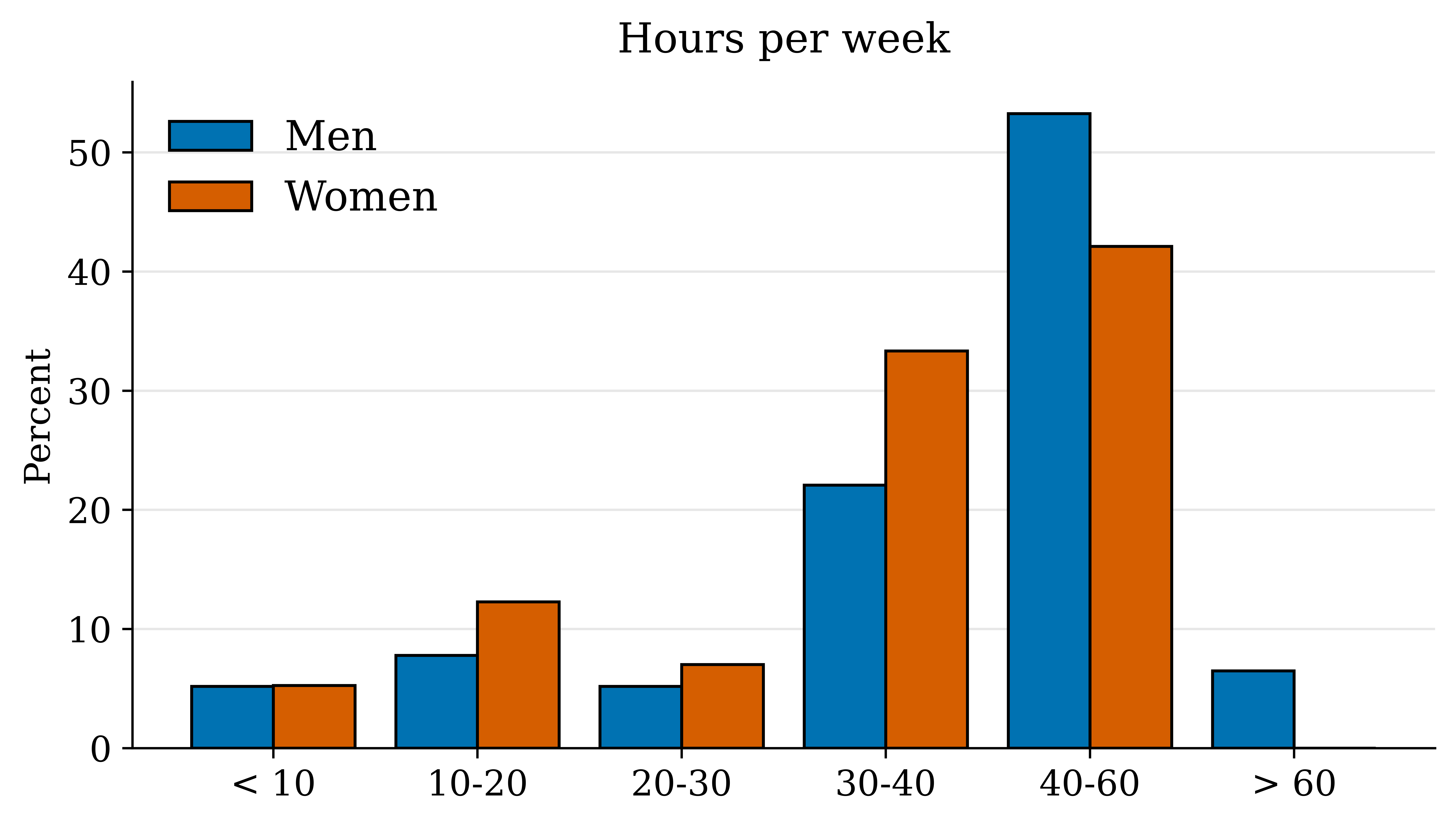}
    \includegraphics[width=0.4\linewidth]{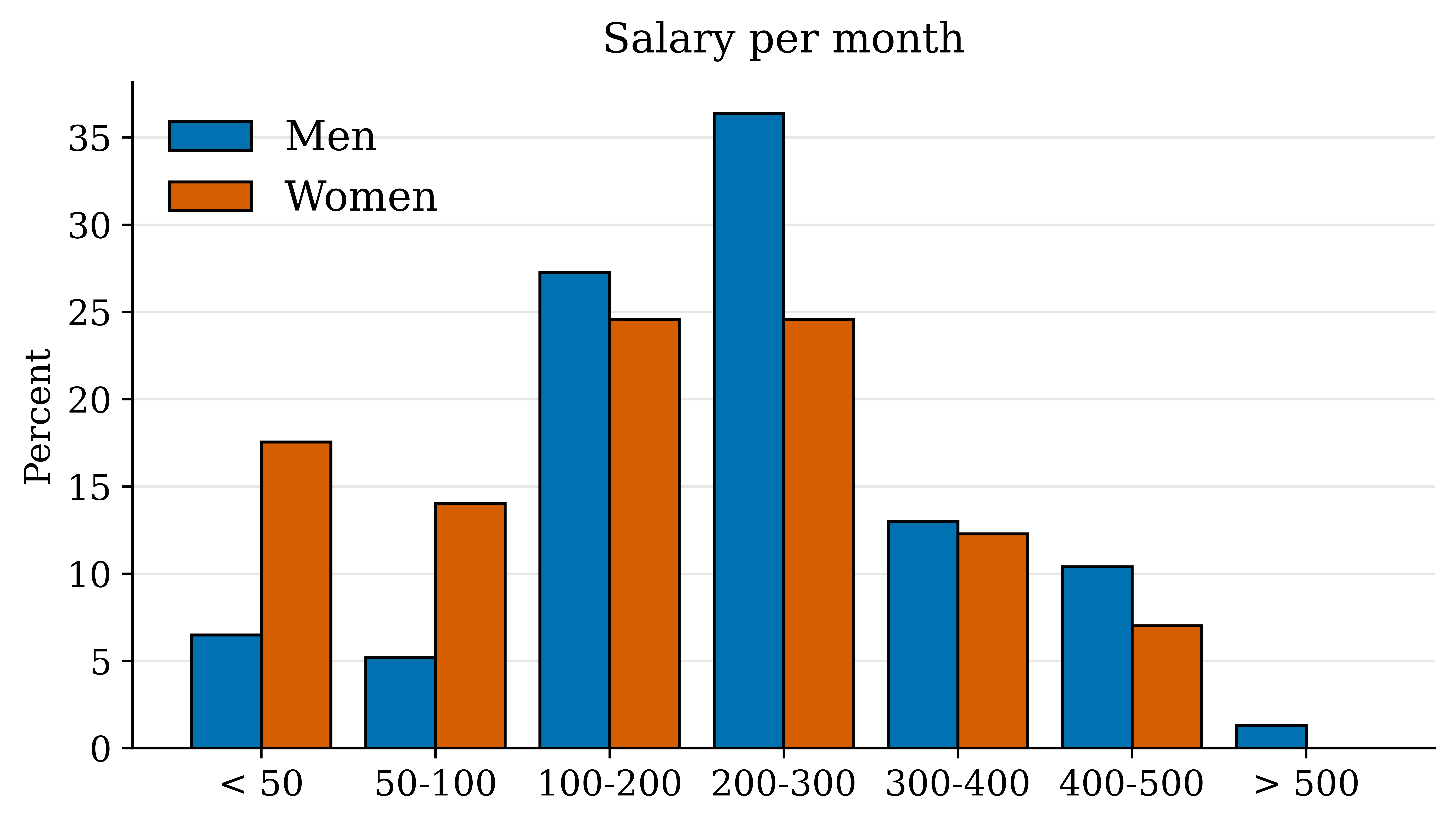}
\end{minipage}

\begin{minipage}{0.89\textwidth}
    \centering
    \includegraphics[width=0.32\linewidth]{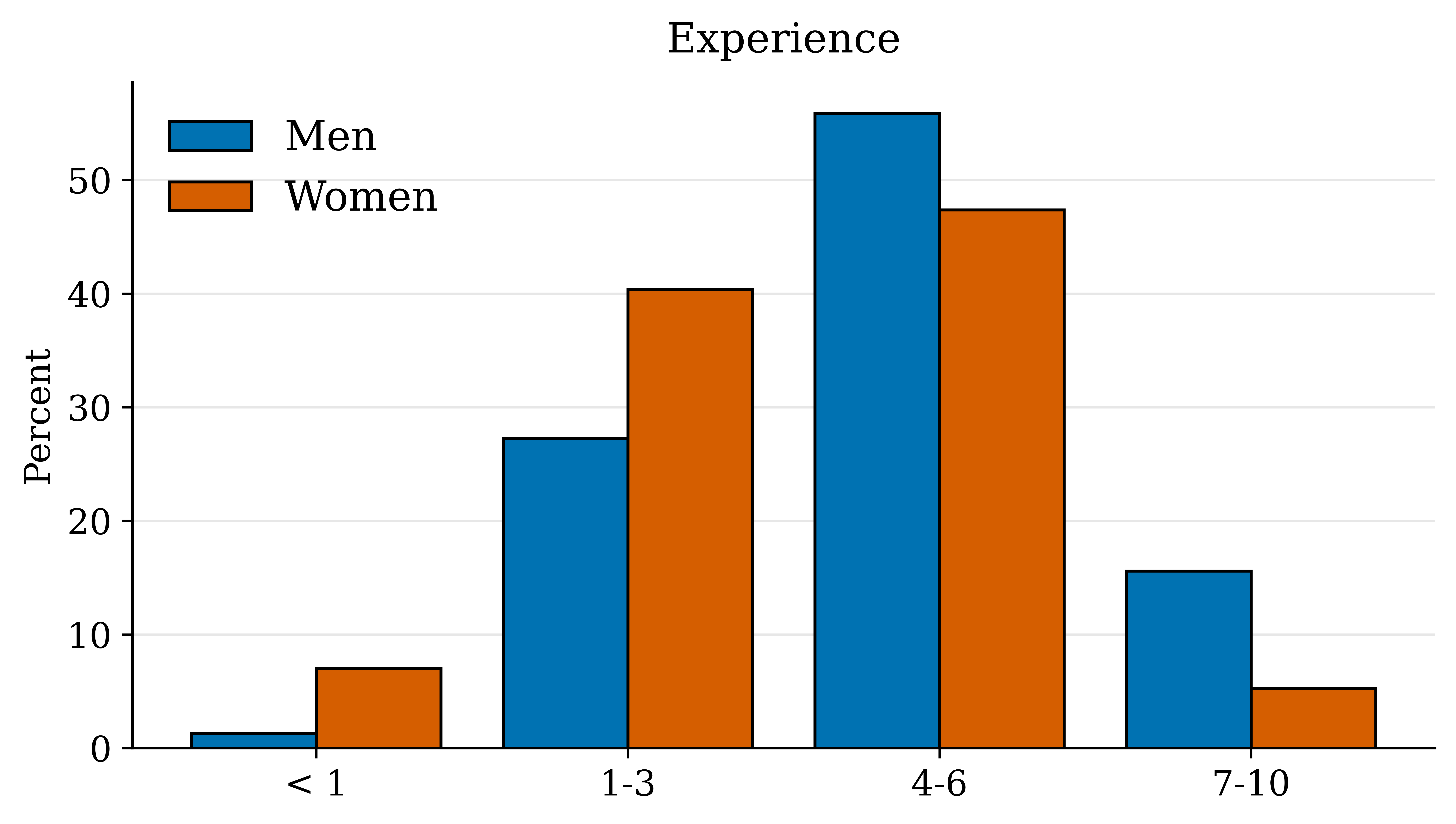}
    \includegraphics[width=0.32\linewidth]{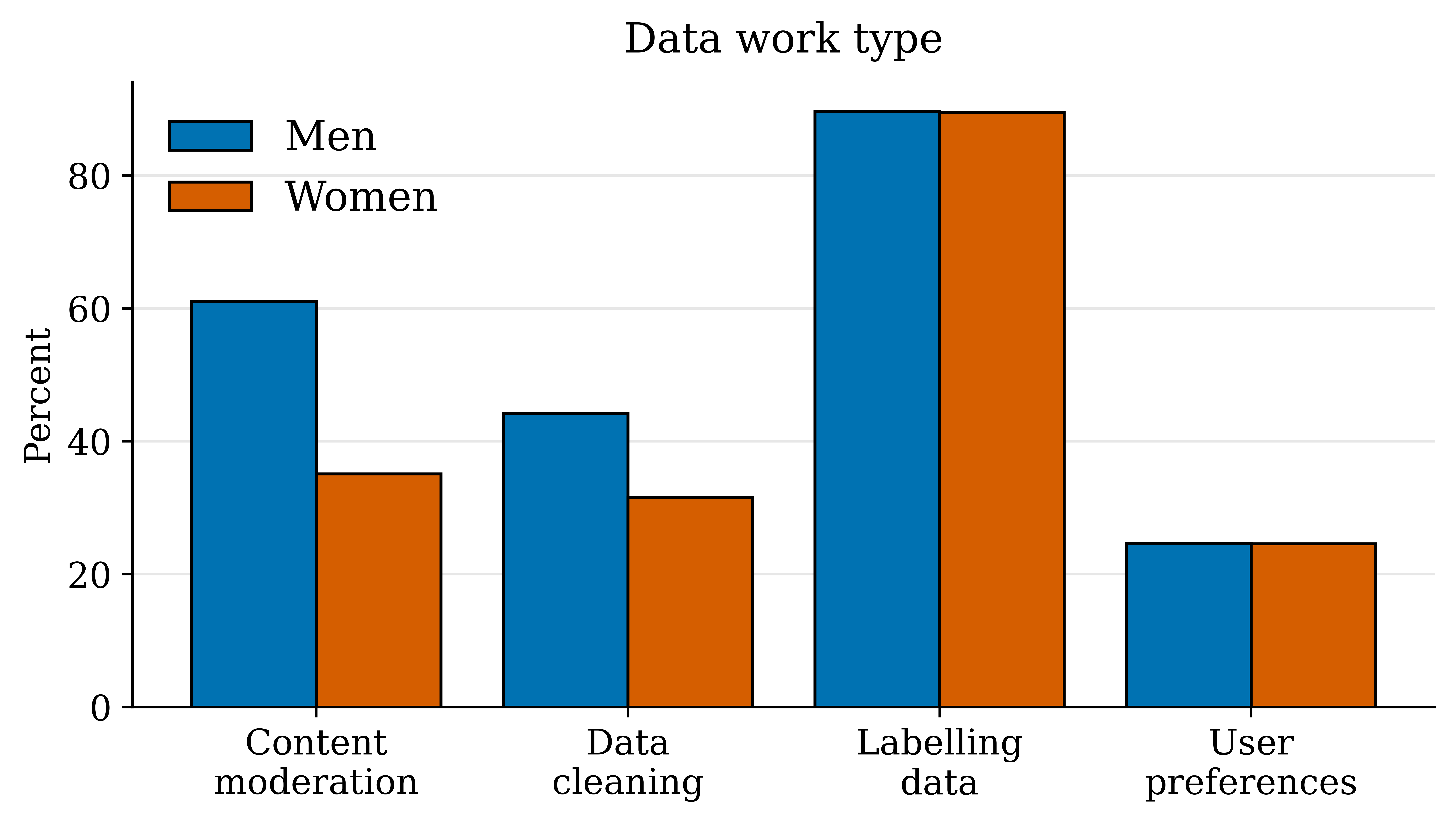}
    \includegraphics[width=0.32\linewidth]{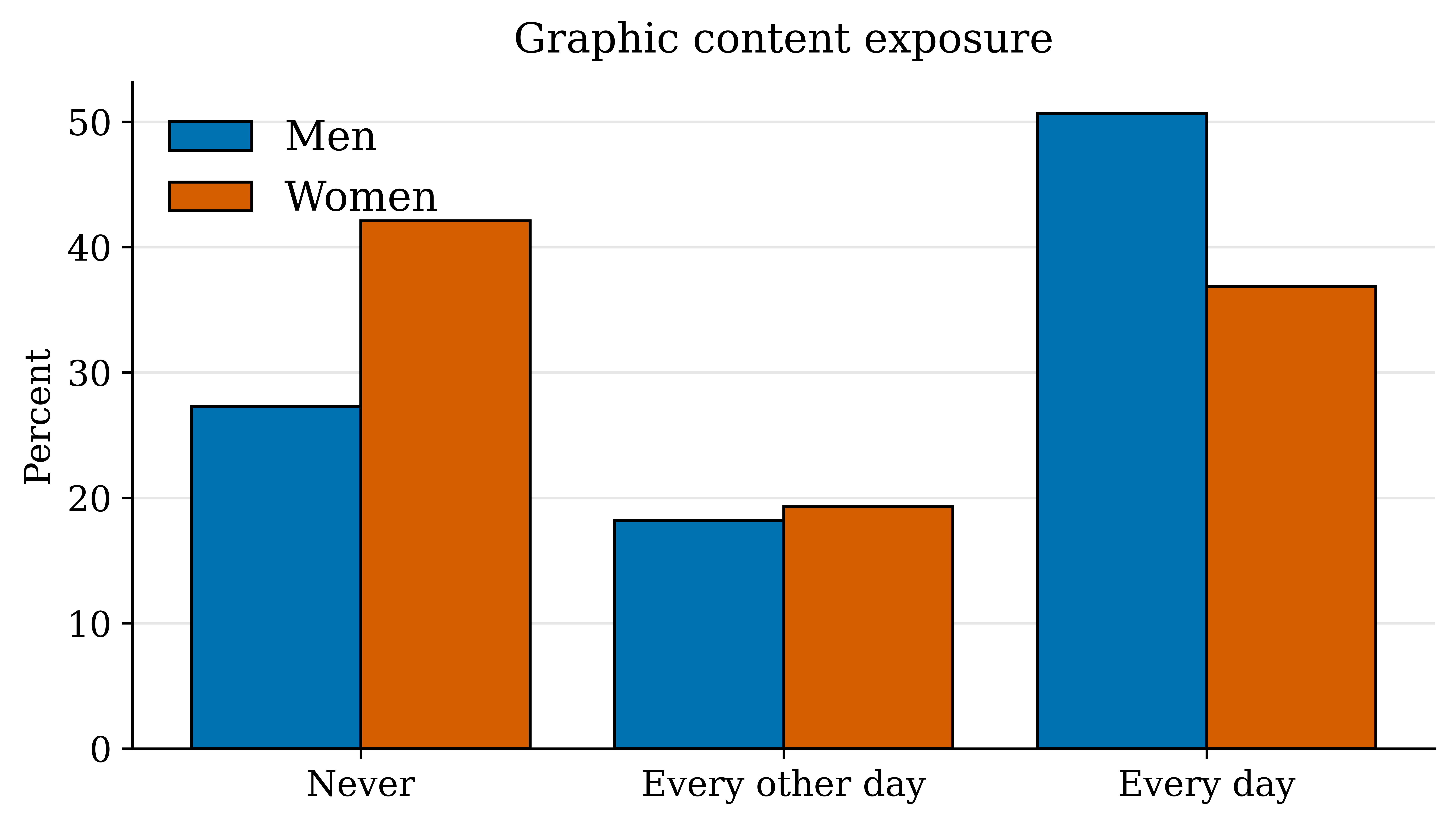}
\end{minipage}
    \caption{Gender-disaggregated distributions of weekly working hours, monthly salary, experience, data work types, and exposure to graphic content derived from the anonymous responses to the questionnaire from 134 data workers.
    }
    \label{fig:gender_distribution}
    \Description{Gender-disaggregated distributions across working conditions in data work. Men are more concentrated in longer working hours, higher salary bands, more experience, and higher exposure to graphic content, compared to women. Task distributions show that men are more frequently engaged in content moderation and data cleaning, while both genders are similarly represented in data labelling and user preference tasks.}
\end{figure}

Gender-disaggregated results in Figure~\ref{fig:gender_distribution} reveal additional disparities. 
Women generally report fewer years of experience, shorter work hours, and lower salaries.
Men report more frequently engaging in content moderation and greater exposure to graphic material.  

{\bf Representation in Datasets.} While hyper-datafication increases the volume of data available, this does not necessarily translate into better representation of the world. We use the datasets stored on Hugging Face to investigate this further.
\looseness=-1

Figure~\ref{fig:datasets_and_data_over_time_bars} (Left and Centre) shows that text data is a large fraction of the diverse data modalities on Hugging Face. Using these text datasets, we first examine the representation of global languages. Among 57,484 datasets with language annotations, we identify 7,934 distinct languages. Most languages (96\%) appear only within multilingual datasets and typically in a small number of repositories. The remaining 4\% correspond to 306 languages, of which 97 occur only in a single dataset. This indicates that many languages exist in the margins of the dataset ecosystem rather than as sustained data sources. 

\begin{figure}[t]
    \centering
    \includegraphics[width=\linewidth]{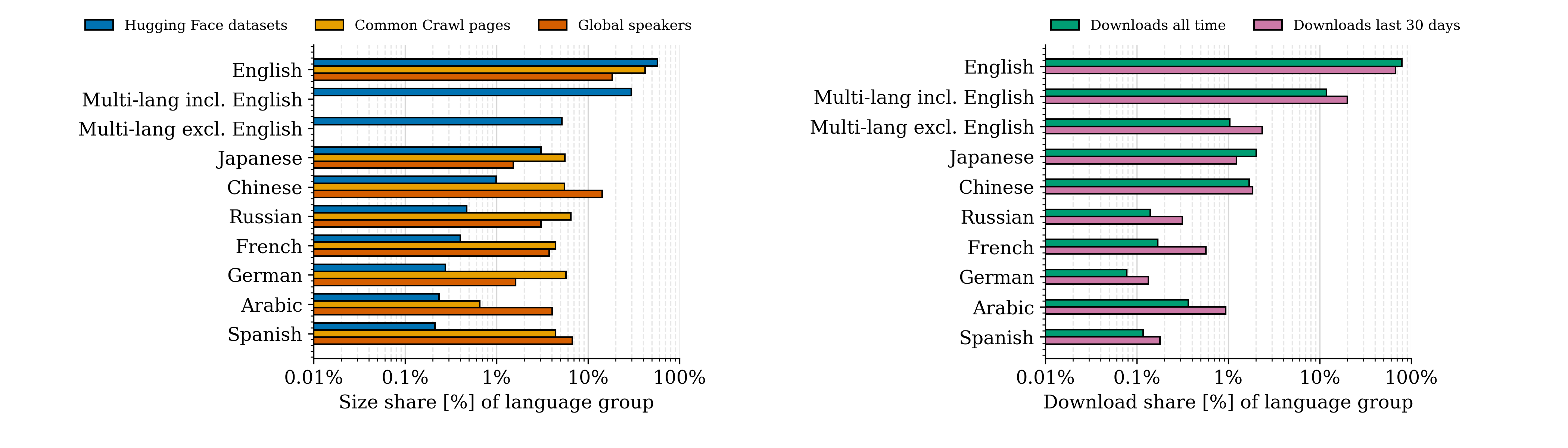}
    \caption{Representation and demand for the ten largest language groups on the Hugging Face Hub. Left: A depiction of each group’s share of dataset volume with its share of Common Crawl pages and global speakers. Right: All-time and recent downloads shares on Hugging Face. Note both horizontal axes are in logarithmic scale.
    }
    \label{fig:language_groups}
    \Description{Representation and demand for the ten largest language groups on the Hugging Face Hub. The left panel compares each group’s share of total dataset size on the Hugging Face Hub with its share of Common Crawl pages and global speakers. English and English-inclusive multilingual datasets dominate the available data, while many widely spoken languages occupy only a small fraction of total dataset volume. The right panel shows the distribution of all-time and last-30-days downloads, where English again accounts for the majority. Recent download shares for non-English languages exceed their all-time shares in most cases, but remain small in absolute terms, indicating only limited shifts in demand.}
\end{figure}

Figure~\ref{fig:language_groups} (Left) shows the representation for the ten largest language groups (in terms of volume share) on the Hugging Face Hub. It compares each group’s share of total dataset size on the Hugging Face Hub with its share of Common Crawl pages and global speakers. English represents a larger share of the dataset volume (57\%) than its share of Common Crawl pages (42\%) and more than three times its share of global speakers (18\%), as shown in Table~\ref{tab:language_distribution}. Multilingual datasets that contain English add another 30\% to this dominance.  
Several widely spoken languages, including Chinese, Arabic, and Spanish, remain substantially under-represented (see Appendix~\ref{app:HF_languages}). Appendix~\ref{app:broadband} provides contextual evidence that these imbalances reflect broader asymmetries in global digital participation and traffic.

Figure~\ref{fig:language_groups} (Right) shows the distribution of downloads in November 2025 and all-time, where English again accounts for the majority. November 2025 download shares for non-English languages exceed their all-time shares in most cases, but remain small in absolute terms, indicating only limited shifts in demand. English-only datasets account for 79\% of all-time downloads and 68\% of November 2025 downloads. 
English-inclusive multilingual datasets account for a further 12\% of all-time and 20\% of November 2025 downloads. 
While November 2025 download shares for several non-English languages exceed their historical averages (see Appendix~\ref{app:HF_languages}), these increases remain small in absolute terms and do not meaningfully reduce English dominance.

\subsection{Economic Costs}
\label{sec:econ}
Hyper-datafication requires substantial investment to store, process, and transmit continuously growing volumes of data. These investments include specialised hardware, networking equipment, cooling systems, and energy infrastructure~\cite{stanford2025,raghav2025sustainable-ai}. As a result, hyper-datafication has a direct and visible economic footprint in the rapid expansion of data centre infrastructure.
\looseness=-1

\begin{figure*}[t]
    \centering
    \includegraphics[height=3cm]{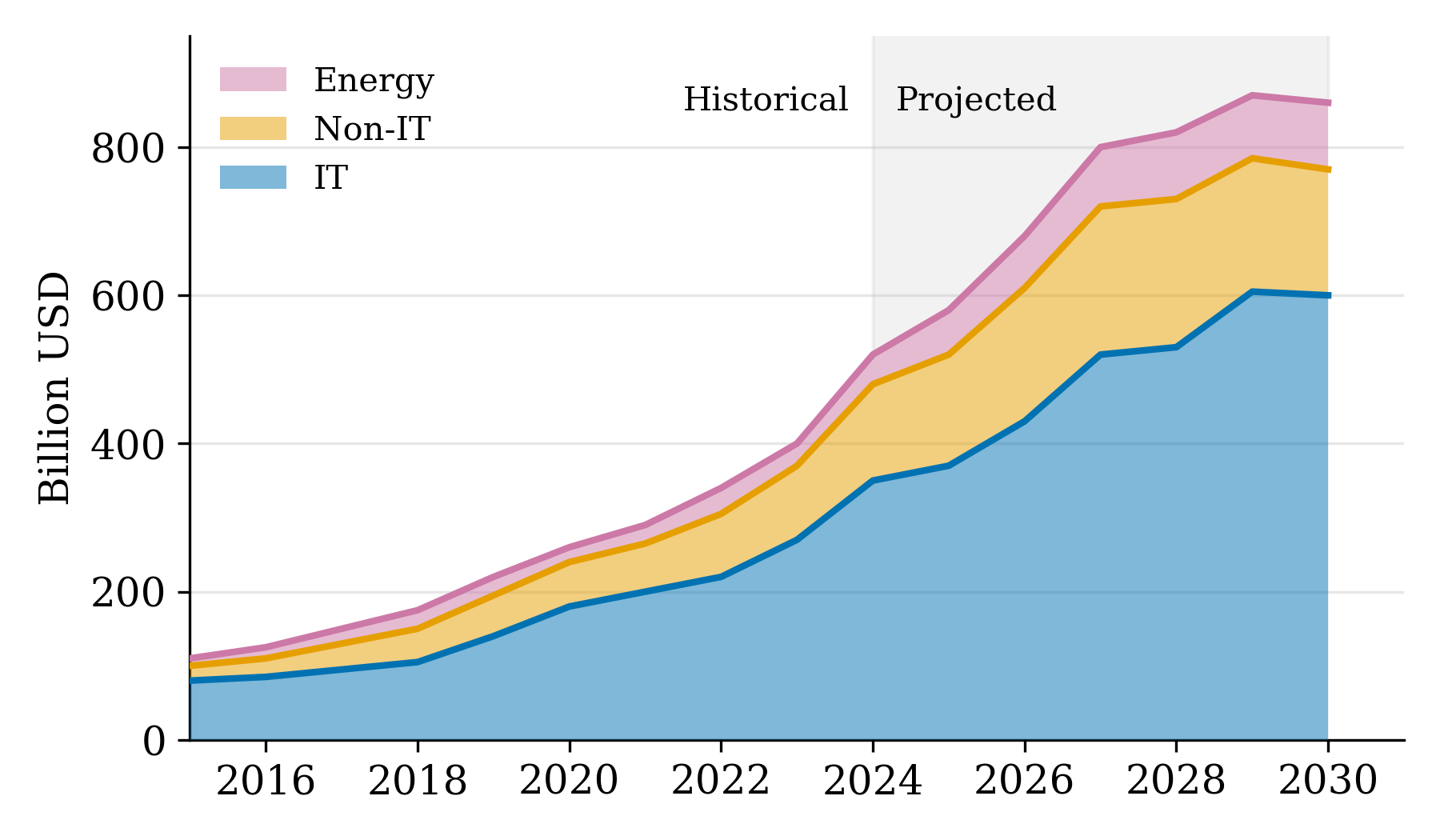}
    \includegraphics[height=3cm]{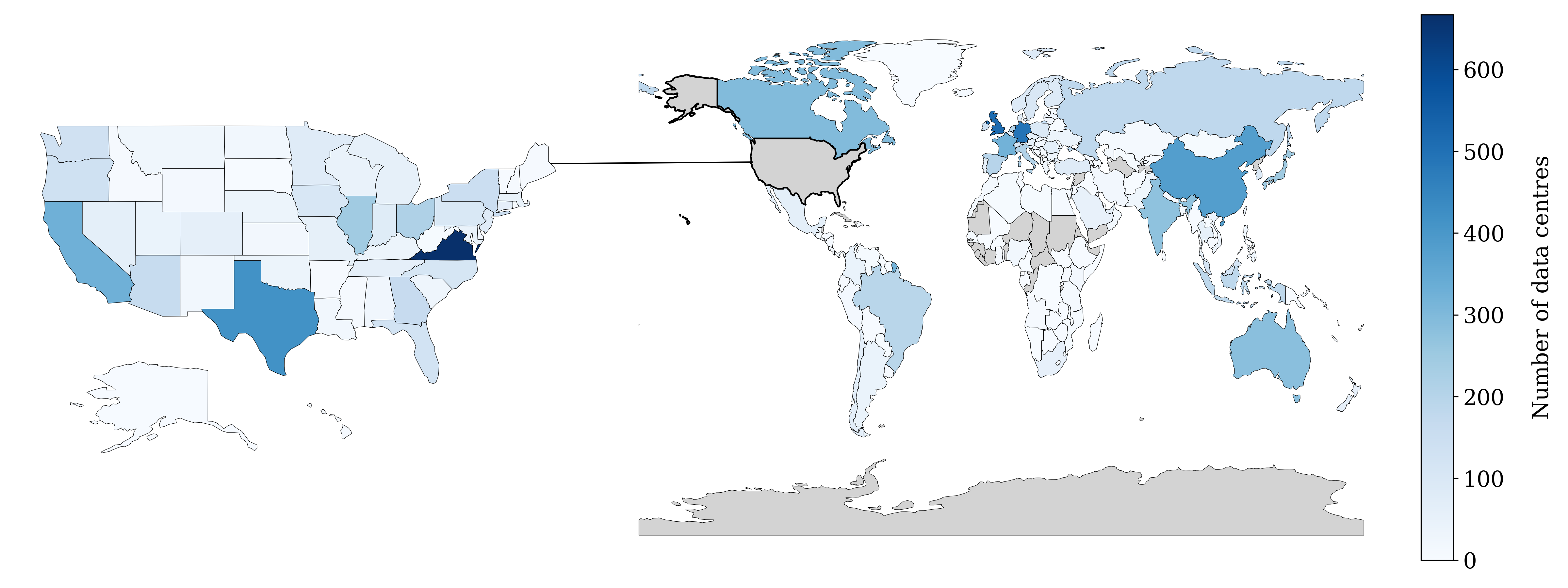}
    \caption{Left: Historical (2015-2024) and projected (2024-2030) global annual investment in data centres in the base case reflecting current regulatory conditions and industry projections as modelled by the IEA~\citep{iea2025}. Right: The world map shows the number of data centres per country. US is highlighted and shown on the left to display state-level variation. Data from Data Center Map~\citep{datacentermap}.}
    \label{fig:datacenters_worldmap}
    \Description{Global scale and geographic concentration of data-centre infrastructure. The figure shows a steady increase in global investment in data centres over time, with projected growth dominated by information technology infrastructure. The accompanying maps show that data centres are unevenly distributed across countries, with a strong concentration in the United States and substantial variation across the states, while many regions host relatively few facilities.}
\end{figure*}

As shown in Figure~\ref{fig:datacenters_worldmap} (Left), global annual investment in data centres has increased more than fivefold over the past decade. These investments are justified by expectations that expanded data capacity will unlock future value through improved AI capabilities. In 2025, global investment in data centres is projected to reach USD 580 billion, exceeding the USD 540 billion invested in global oil supply~\citep{iea2024_weo}. 

This expansion is geographically concentrated, with the US projected to account for more than half of cumulative global data centre investment over the next five years~\citep{iea2025}. 
As shown in Figure~\ref{fig:datacenters_worldmap} (Right), the infrastructure development mirrors this concentration with facilities clustering in Western Europe, China, Australia, Canada, and particularly the US. This spatial pattern only partially aligns with where data is generated and where the social or economic benefits ultimately accrue, as shown by Facebook, whose largest user base is in India with about 384 million users in January 2025, while most of Meta’s infrastructure and revenue are concentrated in the US~\cite{kemp2025facebook}.
\looseness=-1

\section{Discussions}\label{sec:discussion}
{\bf Lack of Transparency.} The environmental costs of hyper-datafication presented in Section~\ref{sec:env} are based on a snapshot of 550,000 datasets on the Hugging Face Hub. Figure~\ref{fig:datasets_and_data_over_time_bars} shows how the scale, size, and diversity of datasets are growing in recent years. Figure~\ref{fig:storage_energy} captures the increase in storage-related operational energy consumption and carbon footprint. However, these estimations do not provide a comprehensive overview of the environmental impact of hyper-datafication. Factors such as the embodied emissions, fresh water to cool data centres, raw materials and minerals used in hardware manufacturing, and the e-waste generated due to hardware disposal are not considered~\cite{falk2025more,raghav2025sustainable-ai}. These broader environmental impacts are difficult to estimate, as acknowledged by works attempting to do this for AI model development~\cite{luccioni2023estimating,wright2023efficiency}. Lack of transparency and openness in reporting environmental costs by vendors, manufacturers, and actors at all stages of the data lifecycle makes comprehensive environmental cost reporting a challenge. 

In addition to the environmental costs, social costs of data, such as labour, are seldom reported and discussed. Results from the data worker questionnaire in Section~\ref{sec:social} highlight these trends where large corporations employ data workers for diverse types of data work. Although longer hours and greater experience are associated with higher earnings, as shown in Figure~\ref{fig:workhours_experience_salary}, salaries remain low relative to the Kenyan national average of USD 540 per month~\citep{CEIC2024KenyaEarnings}. On the one hand, data work creates income opportunities in regions with limited formal employment.\footnote{For example, the youth (15-34 years old) in Kenya face an unemployment rate of 67\%~\citep{fke_youth_employment}.} On the other hand, these same structural conditions enable exploitation~\cite{shitawa2024click}. The availability of surplus labour, combined with limited regulatory oversight, allows data work to be organised through precarious contracts, low wages, and intensive performance monitoring. Economic necessity weakens workers' bargaining power, while task fragmentation and platform-based subcontracting obscure accountability. Transparent reporting of the costs pertaining to data work can give due credit to the data workers who remain unseen in the discourse surrounding frontier AI. Additionally, the credit and visibility can help them secure better work conditions and bargaining rights.
\looseness=-1

Although enforcing such reporting is not straightforward, it could be encouraged through incentive structures that offer companies strategic recognition for disclosing data-related environmental and social impacts. For example, certification of sustainable data practices could reward companies for comprehensive reporting. This could frame transparency as a competitive advantage rather than merely a regulatory burden. 

{\bf Material Costs of Data and Resource-Awareness.} The narrative of data ``infiniteness'' behind hyper-datafication is based on the view that data has a negligible cost, and ignores the material footprint of data. In Section~\ref{sec:env}, we presented concrete estimations of the energy consumption and carbon footprint of data using Hugging Face datasets. Extending these estimations to all data currently used for AI development will yield even higher material costs. 
Measuring the cost of creating, storing, and processing datasets using existing tools like Carbontracker\footnote{\url{https://carbontracker.info/}}~\cite{anthony2020carbontracker} or CodeCarbon\footnote{\url{https://codecarbon.io/}} and {reporting the carbon emissions in standardised datasheets} as proposed by \citet{gebru2021datasheets} are good starting points. Additionally, attaching environmental costs at each step as meta-data across the data life-cycle, as suggested by \citet{mersy2024toward} for carbon-provenance-based AI, can clarify the assessment of the environmental sustainability of data. 
\looseness=-1

{\bf Data Provenance and Ownership under Data Monopolisation.} Another factor that is propelling hyper-datafication is the belief that more data necessarily creates more value, which also drives data concentration. Only a small number of firms, based in a handful of countries, possess the capital and infrastructure required to collect, store, and process data at hyper-scale. Figure~\ref{fig:datacenters_worldmap} shows this spatial disparity, where vast regions of the world have few data centres, whereas the majority of the global data centres are located in the US. 
Such concentration creates high barriers to entry and reinforces monopolistic dynamics~\cite{mejias2024data}. 

The concentration of data accumulation enables large-scale value extraction from user-generated and third-party data without corresponding compensation or meaningful control for data producers. Individuals and organisations contribute data under conditions of limited transparency, while platforms and the countries they belong to capture the resulting economic surplus \cite{jia2025a}. The outcome is a persistent structural asymmetry in value appropriation \cite{jia2025a}. Evaluating how value is created from data requires {\em comprehensive data provenance}, which is not straightforward with the frontier AI systems~\cite{longpre2024large}. This is also closely tied to the question of {\em data ownership}, in which big tech platforms exploit data created by users to create value for themselves through frontier AI systems, manifesting extreme forms of appropriation~\cite{zuboff2023age}. Protecting people’s rights in frontier AI systems, such as generative AI models, relies on data-provenance mechanisms and expanded legal protections, such as granting individuals copyright over their own identity. Recent proposals, including the Danish model of giving the copyright of their identity to people, illustrate how data and identity rights could be formally recognised in this way~\cite{bryant2025denmark}.
\looseness=-1

{\bf Data Frugality to Counter Hyper-Datafication.} 
Hyper-Datafication promotes a \emph{data abundance mindset} in which ever-larger datasets are treated as the primary driver of model progress, based on empirical scaling laws linking model performance to data and compute \citep{kaplan2020scaling, hoffmann2022training}.
Yet, our findings, evidenced by the storage-related energy use (Section~\ref{sec:env}), data labour conditions (Section~\ref{sec:social}), and infrastructure expansion (Section~\ref{sec:econ}), show that this logic systematically externalises social and environmental costs: the storage and curation of ever-growing datasets impose increased labour burdens and raise the energy and carbon costs of data storage, curation, and model training.
Those costs grow faster than the marginal performance gains they produce \citep{Bakhtiarifard_ECNAS_2024}.

However, a frugal alternative exists, which builds on the well-established fact that not all data is equally informative \cite{katharopoulos2018not}. 
Techniques such as representative subset selection and coresets identify a small, representative, and potentially reweighted subset of data that approximately preserves the learning dynamics of a full dataset.
This allows models to achieve comparable performance with substantially less data and computation \citep{killamsetty2021grad,sorscher2022beyond,muennighoff2023scaling} and reduced environmental cost \citep{wilson2026stop}.
Despite these benefits, these approaches often require one-time upfront construction effort, which is amortised when the resulting subset is sufficiently small or used multiple times, for example, to tune model hyperparameters.
Nevertheless, under current economic incentives, there is little motivation for companies to adopt such approaches.
Data accumulation increases not only technical capacity but also strategic control and market valuation, encouraging data hoarding even when its utility is low.
Finally, companies may simply retain data due to a ``fear of missing out'' on valuable insights or future utility \citep{gormley2012data,zuboff2023age}.

For this reason, data frugality cannot be left to voluntary optimisation.
It must be enforced as a structural constraint on hyper-datafication.
A {\em corporate data tax} could internalise the social and environmental costs of data accumulation, correcting a market failure in which companies benefit from scale while communities and workers bear the burdens. 
Thus, the tax could discourage wasteful hoarding while generating revenue to compensate data generators and data workers, and increase transparency around data ownership and use.
A related proposal by~\citet{irwin2026tokentaxesmitigatingagis} suggests a data ``token tax'' on AI inference to capture value where AI is used rather than where the infrastructure is hosted.
Tax-based instruments are, however, most often jurisdiction-specific and risk displacing activity to data tax havens. Effective implementation therefore requires frameworks with extraterritorial reach and binding legal force, such as the EU AI Act, which could serve as a basis for introducing data cost-related accountability mechanisms.
Frugality, in this sense, is not merely an efficiency principle but a corrective action, which must be backed by enforceable policy to meaningfully change the extractive political economy of data~\cite{pasquinelli2023eye}. 

{\bf Recommendations to Mitigate Costs of Hyper-datafication.} 
We have presented several critiques of hyper-datafication in this section substantiated by evidence from Section~\ref{sec:costs}. We synthesise these arguments into a set of recommendations, ranging from concrete technical measures to longer-term visions for the future, aimed at multiple stakeholders. 

\begin{tcolorbox}[colback=blue!10, colframe=blue!10]
{\bf Data PROOFS Recommendations.}
\begin{itemize}
    \item {\bf Provenance:} Build data provenance into frontier AI systems, for monitoring data quality and attributing due credit to data producers. This requires the development of new tools and regulations. 
    \item {\bf Resource-awareness:} Measure and report the environmental and labour costs of data to counter the narrative about data ``infiniteness'' and materially ground the bits.
    \item {\bf Ownership:} Unless explicitly agreed, the data ownership should--by default--belong to users and not platforms. This must be encoded as digital citizen rights. 
    \item {\bf Openness:} Transparent reporting of data costs. Availability of data used for frontier AI development, so that researchers and policy makers can use it for inquiry. 
    \item {\bf Frugality:} Data frugality should be the guiding design principle instead of data abundance. Incentivised as a corporate data tax, which can then be used to compensate data owners and data workers. 
    \item {\bf Standards:} Develop measurement and reporting standards that can facilitate sustainability provenance by embedding environmental, social, and monetary costs as meta-data along the data life-cycle.
\end{itemize}
\end{tcolorbox}

In addition to the Data PROOFS recommendations, {\bf collective action} is indispensable to counter the consequences of hyper-datafication and to improve the sustainability of data in frontier AI. Community resistance is growing; in the US, more than 140 activist groups across 24 states are mobilising against new data centre construction and expansion~\citep{datacenterwatch2025_report}. Over the past two years, these groups have collectively stalled or cancelled proposed projects worth USD 162 billion. Citizen-led protests against hyper-scale data centres are appearing in Mexico, Spain, the Netherlands, India, and other parts of the world~\cite{mozr2025mexico}. 
This trend highlights how the environmental, social, and economic costs of data centre development are becoming increasingly visible, and how local communities are contesting the uneven distribution of benefits and harms associated with hyper-scale AI development.  
\looseness=-1

{\bf Limitations.} We used datasets from Hugging Face Hub as it is one of the few platforms where dataset metadata is available in a standardised format. Extrapolating our arguments to broader data trends in the AI domain is a limitation of our work. However, Hugging Face Hub has evolved into a reliable platform for frontier AI data and models, and serves as a reasonable sample that captures the broader trends and has been used in other works~\cite{yang2024navigating,castano2024analyzing,luccioni2024power}.
Thus, our estimates can be seen as lower bounds on true data costs.
Particularly, as we have noted in Section~\ref{sec:social}, the language representation on the Hugging Face Hub is highly uneven. 
It does not reflect the global distribution, nor does it mirror global patterns of digital activity. 
This should not be interpreted as a failure of a single platform. 
Rather, it reflects broader structural dynamics in how data is produced and used across the AI ecosystem. 

Data work is carried out globally, but our questionnaire was limited to workers in Kenya. We have refrained from making generalised claims and tried to limit the scope of labour discussions to the conditions in Kenya. 
\looseness=-1

Likewise, some of the economic costs are primarily US-focused. This is again due to the availability of data sources. Furthermore, it reflects the dominance of a few regions, such as the US, which host the majority of global infrastructure for frontier AI development. Finally, while the geographic concentration of data centres captures their presence, it does not reflect their purpose, scale or utilisation.

\section{Conclusions}
\label{sec:conclusions}

The push for hyper-datafication is tied to the objective of modelling reality from data. 
In this pursuit, we are progressing from an era in which models relied mainly on passively accumulated data to one in which the world is actively reshaped to generate more data. Sensors, platforms, and infrastructures increasingly optimise for continuous data extraction. While hyper-datafication increases the amount of available data, it does not necessarily yield informative and useful data. Furthermore, hyper-datafication does not alleviate existing problems with data-driven models but can aggravate them.
\looseness=-1

We have presented evidence from multiple sources that bring forth the sustainability costs of hyper-datafication in frontier AI. 
The results presented in this work invite us to take a step back and reflect on the pursuit of frontier AI on the back of internet-scale data. 
Hyper-datafication is framed as an inevitable route to more capable and general AI systems. 
Our findings suggest that this is not a neutral form of progress. 
Instead, it redistributes benefits and burdens unevenly, while deepening already existing asymmetries in representation and power. 
We hope the Data PROOFS recommendations serve as inspiration for improving the overall sustainability of data in frontier AI.

\begin{acks}
SW and RS acknowledge funding from the Independent Research Fund Denmark (DFF) under grant agreement No. 4307-00143B. 
RS and JK acknowledge funding from the European Union's Horizon Europe Research and Innovation Action programme (HORIZON-CL4-2021-HUMAN-01) under grant agreement No. 101070408, project SustainML (Application Aware, Life-Cycle Oriented Model-Hardware Co-Design Framework for Sustainable, Energy Efficient ML Systems). 
RS further acknowledges funding from the European Union's Horizon Europe Research and Innovation Action programme under grant agreements No. 101070284 and No. 101189771. 
ED acknowledges funding from the Novo Nordisk Foundation under grant agreement No. NNF24OC0090977.
SM acknowledges funding from the Danish Data Science Academy (DDSA) through DDSA Visit Grant No. 2025-5673.
The authors thank the data workers who participated in the anonymous questionnaire for their time and contributions, and members of \href{https://saintslab.github.io/}{SAINTS Lab} for valuable discussions.
The authors also thank the anonymous reviewers and the area chair at FAccT for their valuable feedback. 
\end{acks}

\section*{Author Contributions}
Contributions follow the CRediT taxonomy\footnote{https://credit.niso.org/}.
{\bf SW:} Conceptualization, Data curation, Formal analysis, Investigation, Methodology, Project administration, Resources, Software, Validation, Visualization, Writing – original draft, Writing – review \& editing. 
{\bf SM:} Methodology, Validation, Writing – original draft, Writing – review \& editing. 
{\bf MO:} Investigation, Resources. 
{\bf ED:} Conceptualization, Supervision, Validation, Writing – original draft, Writing – review \& editing. 
{\bf JK:} Conceptualization, Data curation, Methodology, Resources, Validation, Writing – original draft, Writing – review \& editing. 
{\bf RS:} Conceptualization, Data curation, Investigation, Methodology, Project administration, Resources, Supervision, Validation, Visualization, Writing – original draft, Writing – review \& editing.

\section*{Generative AI Usage Statement}
GitHub Copilot and ChatGPT version 5.1 were used to support programming tasks, including the development of scripts for Hugging Face metadata extraction and data visualisation. Google Gemini, ChatGPT versions 5.1 and 5.2 were used to identify relevant academic and policy sources, assist with table generation, and to edit and refine language and grammar in selected sections of the manuscript.

\balance
\bibliographystyle{ACM-Reference-Format}
\bibliography{refs}

\appendix
\clearpage \newpage
\section*{Appendix}

\section{Hugging Face Datasets Metadata} \label{app:HF_metadata}

\subsection{Metadata Coverage} \label{app:metadata_coverage}
Table~\ref{tab:HF_metadata} provides an overview of all metadata attributes and coverage. 

\begin{table}[h!]
\centering
\footnotesize
\caption{Overview of metadata attributes. The final sample includes 554,300 datasets.}
\label{tab:HF_metadata}
\begin{tabular}{llr}
\toprule
\textbf{Attribute} & \textbf{Description} & \textbf{Coverage [\%]} \\
\midrule
\texttt{id} & Unique dataset identifier on the Hub  & 100.0 \\
\texttt{created\_at} & Timestamp of repository creation & 100.0 \\
\texttt{last\_modified} & Timestamp of the most recent commit & 100.0 \\
\texttt{downloads\_all\_time} & Total number of downloads since creation & 100.0 \\
\texttt{downloads\_30d} & Number of downloads in the last 30 days & 100.0 \\
\texttt{used\_storage} & Estimated storage used by the dataset on the Hub (bytes) & 91.0 \\
\texttt{dataset\_size} & Local storage footprint of the dataset (bytes) & 74.7 \\
\texttt{region} & Reported hosting region (US or EU) & 99.8 \\
\texttt{modality}$^*$ & Data modality  & 78.3 \\
\texttt{task}$^*$ & High-level task category  & 12.9 \\
\texttt{sub-task}$^*$ & Specific task   & 12.9 \\
\texttt{language}$^*$ & Languages represented in the dataset (ISO-based codes) & 10.2 \\
\bottomrule
\end{tabular}%

\footnotesize{$^*$ indicates that the attribute allows multiple labels.}
\end{table}

\subsection{Modalities and Tasks} \label{app:metadata_modalities}
Modalities include text, image, tabular, video, audio, time-series, 3d, document, and geospatial. The tasks include audio-speech (AS), computer vision (CV), multimodal (MM), natural language processing (NLP), reinforcement learning (RL), and tabular (TAB). 

Datasets without a declared modality account for 22\% of all datasets but represent 51\% of total storage volume, even though for 41\% of them the dataset size could not be extracted. The dominant grey area in Figure~\ref{fig:datasets_and_data_over_time_bars} (Centre) therefore corresponds to only 14\% of all datasets. 

\begin{figure}[h]
    \centering
    \includegraphics[width=0.8\linewidth]{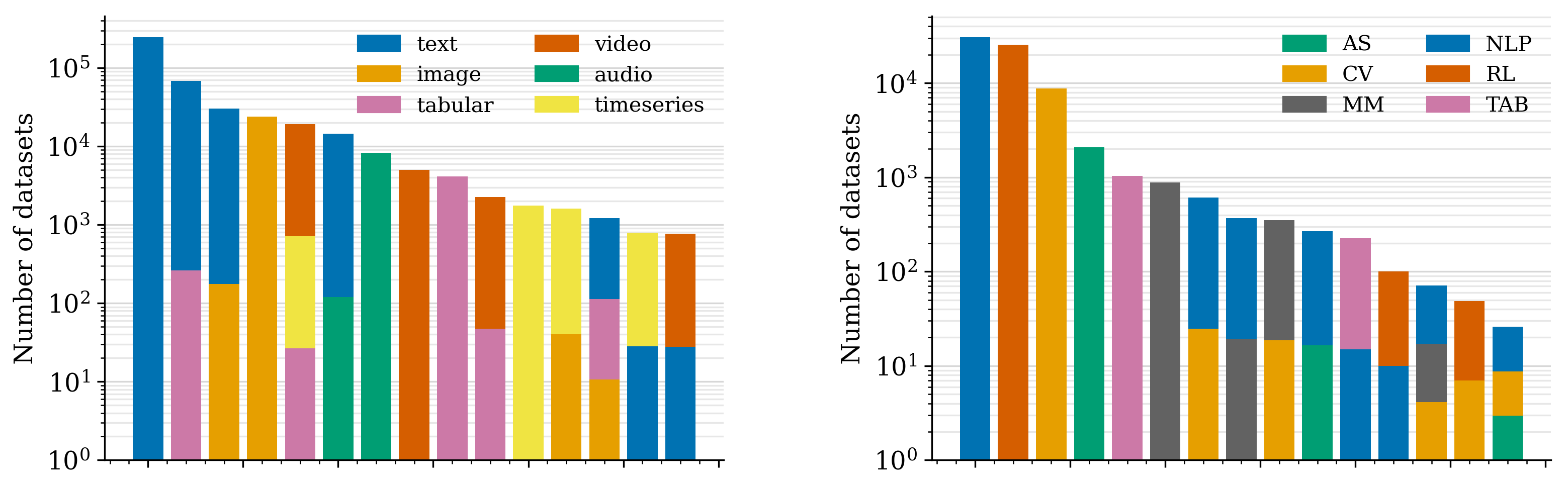}
    \caption{Distribution of dataset modalities and task categories on the Hugging Face Hub. Left: The fifteen most common dataset modality combinations on a logarithmic scale. Right: The fifteen most common dataset task combinations on a logarithmic scale.}
    \label{fig:modalities_and_tasks}
    \Description{Distribution of dataset modalities and task categories on the Hugging Face Hub. The figure shows that text-based datasets dominate modality combinations by a large margin. Task categories are somewhat more even, with natural language processing and reinforcement learning datasets most common, followed by computer vision, and substantially fewer datasets in multi-modal, tabular, and audio-related tasks.}
\end{figure}

\begin{figure}[h]
    \centering
    \includegraphics[width=0.8\linewidth]{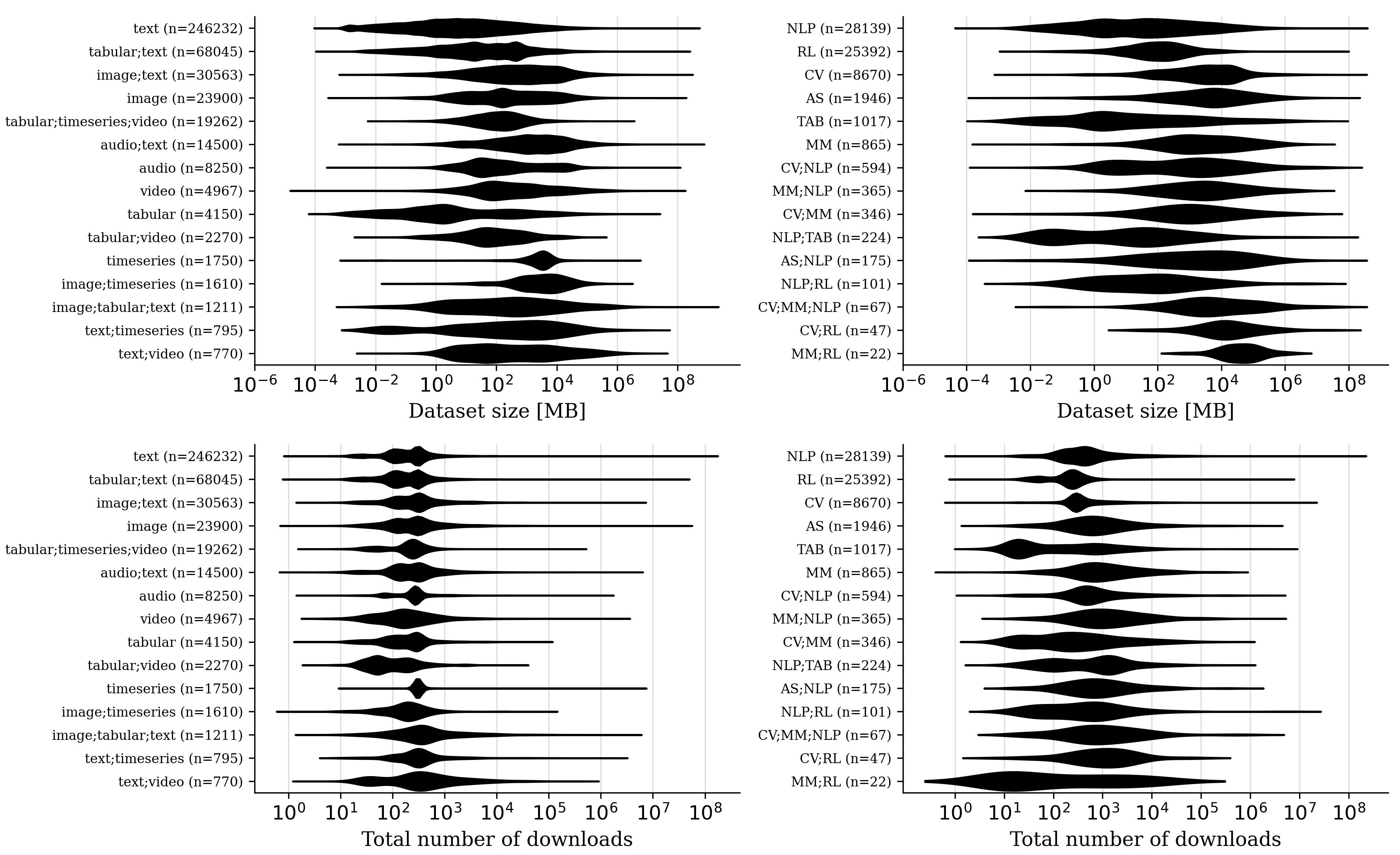}
    \caption{Distribution of dataset sizes and downloads on the Hugging Face Hub by modality and task. Left: Violin plots of dataset sizes (top) and total downloads (bottom) for the fifteen most common modality combinations. Right: Violin plots of dataset sizes (top) and total downloads (bottom) for the fifteen most common task combinations.}
    \label{fig:violin_plots}
    \Description{Distributions of dataset sizes and download counts across modality and task combinations on the Hugging Face Hub. Dataset sizes span several orders of magnitude across all combinations including both modalities and tasks, reaching differences of up to thirteen orders of magnitude. Download counts show similarly extreme dispersion, spanning up to eight orders of magnitude. }
\end{figure}

Figure~\ref{fig:modalities_and_tasks} shows the distribution of modality and task combinations, with text datasets dominating the modality landscape. Figure~\ref{fig:violin_plots} shows the distribution of dataset sizes and total downloads across modalities and tasks.

\subsection{Language and Download Distribution for Top Ten Languages} \label{app:HF_languages}
To support Figure~\ref{fig:language_groups}, Table~\ref{tab:language_distribution} reports language shares on the Hugging Face Hub by dataset volume, compared with Common Crawl page shares and global speaker populations. Table~\ref{tab:language_downloads} reports the corresponding shares of total downloads and downloads in November 2025 corresponding to the 30 days preceding data collection.

\begin{table}[H]
\centering
\footnotesize
\caption{Language distribution across Hugging Face (HF) datasets, Common Crawl (CC) pages, and global speaker populations for the top ten language groups by volume.}
\label{tab:language_distribution}
\begin{tabular}{lrrr}
\toprule
\textbf{Language group} & \textbf{HF dataset size (\%)} & \textbf{CC pages (\%)} & \textbf{Speakers (\%)} \\
\midrule
English & 57.4 & 42.1 & 18.5 \\
Multi-lang incl. English & 29.8 & -- & -- \\
Multi-lang excl. English & 5.2 & -- & -- \\
Japanese & 3.1 & 5.6 & 1.5 \\
Chinese & 1.0 & 5.6 & 14.3 \\
Russian & 0.5 & 6.5 & 3.1 \\
French & 0.4 & 4.4 & 3.8 \\
German & 0.3 & 5.7 & 1.6 \\
Arabic & 0.2 & 0.7 & 4.1 \\
Spanish & 0.2 & 4.4 & 6.8 \\
\bottomrule
\end{tabular}
\end{table}

\begin{table}[H]
\centering
\footnotesize
\caption{Download distribution across Hugging Face datasets for the top ten language groups by volume.}
\label{tab:language_downloads}
\begin{tabular}{lrr}
\toprule
\textbf{Language group} & \textbf{Downloads, all time (\%)} & \textbf{Downloads, last 30 days (\%)} \\
\midrule
English & 79.4 & 67.6 \\
Multi-lang incl. English & 11.8 & 20.1 \\
Multi-lang excl. English & 1.0 & 2.4 \\
Japanese & 2.0 & 1.2 \\
Chinese & 1.7 & 1.9 \\
Russian & 0.1 & 0.3 \\
French & 0.2 & 0.6 \\
German & 0.1 & 0.1 \\
Arabic & 0.4 & 0.9 \\
Spanish & 0.1 & 0.2 \\
\bottomrule
\end{tabular}
\end{table}

\section{Questionnaire for Data Workers} \label{app:questionnaire}
The questionnaire consisted of ten questions covering demographics, working conditions, and exposure to graphic content. All questions were administered in English. 
The questionnaire can be seen below.

\paragraph{Employment and Work Characteristics}

\begin{itemize}
\item \textbf{What type of data work do you typically do? (multiple selections allowed)}
\begin{itemize}
\item Labelling data (images, text, etc.)
\item Content moderation
\item Verification of user preferences (e.g. in chatbots)
\item Data cleaning
\item None of the above
\item Other (free-text)
\end{itemize}
\item \textbf{How often are you exposed to graphic content (violence, self-harm, hate speech, sexual abuse etc.) as part of your data work?}
\begin{itemize}
    \item Never
    \item 1--2 times a day
    \item Several times each day
    \item Every other day
    \item Other (free-text)
\end{itemize}

\item \textbf{Have you directly worked (not via intermediate BPOs or platforms) for any of the following big tech companies? (multiple selections allowed)}
\begin{itemize}
    \item Google
    \item Microsoft
    \item Facebook/Meta
    \item OpenAI
    \item Amazon
    \item None of the above
    \item Other (free-text)
\end{itemize}

\newpage
\item \textbf{How long have you worked as a data worker?}
\begin{itemize}
    \item Less than one year
    \item 1--3 years
    \item 4--6 years
    \item 7--10 years
    \item More than 10 years
\end{itemize}

\item \textbf{How many hours of data work do you typically do per week?}
\begin{itemize}
    \item Less than 10 hours
    \item 10--20 hours
    \item 20--30 hours
    \item 30--40 hours
    \item 40--60 hours
    \item More than 60 hours
\end{itemize}

\item \textbf{What is your average gross monthly salary (before taxes) for doing the data work?}
\begin{itemize}
    \item Less than 50 USD
    \item 50--100 USD
    \item 100--200 USD
    \item 200--300 USD
    \item 300--400 USD
    \item 400--500 USD
    \item More than 500 USD
\end{itemize}
\end{itemize}

\paragraph{Demographics}
\begin{itemize}
\item \textbf{Gender:}
\begin{itemize}
\item Female
\item Male
\item Prefer not to say
\item Other (free-text)
\end{itemize}
\item \textbf{Age bracket:}
\begin{itemize}
    \item Under 20
    \item 20--30
    \item 30--40
    \item Older than 40
\end{itemize}

\item \textbf{Which country are you based in?}
\begin{itemize}
    \item Free-text
\end{itemize}
\end{itemize}

\paragraph{Additional Comments.}
\begin{itemize}
\item{\textbf{Any additional points you would like to add? (optional)}}
\begin{itemize}
    \item Free-text
\end{itemize}
\end{itemize}

\section{Questionnaire Summary Statistics} \label{app:questionnaire_raw_counts}
Responses were preprocessed prior to analysis to ensure consistency across categories. For exposure to graphic content, the questionnaire included four fixed-response options (Never, 1–2 times a day, Several times each day, Every other day). The two daily exposure categories (1–2 times a day and Several times each day) were collapsed into a single category labelled Every day. One free-text response reported ``Everyday'' which was also mapped to Every day.  
A small number of free-text responses (e.g. ``sometimes'', ``once in a while'') were grouped into a single Sometimes category and excluded from further analysis due to low frequency (4/134).  

Free-text responses for type of data work were not included in the analysis. These responses were all unique single mentions (e.g., ``AI training'', ``customer service'', ``3D''). 

For employment relationships with large technology companies, free-text responses were used to clarify indirect work arrangements (e.g., work conducted via business process outsourcing (BPO) firms). All such responses were mapped to the predefined category None of the above, reflecting indirect rather than direct employment.

The preprocessed counts are summarised in Table~\ref{tab:raw_counts_questionnaire}.

{\setlength{\tabcolsep}{3pt}
\begin{table}[h!]
\centering
\footnotesize
\caption{Raw counts for all categorical variables and data work types.}
\label{tab:raw_counts_questionnaire}

\begin{tabular}{@{}
p{1.6cm} r
@{\hspace{0.3cm}\vrule\hspace{0.3cm}}
p{1.9cm} r
@{\hspace{0.3cm}\vrule\hspace{0.3cm}}
p{3.0cm} r
@{\hspace{0.3cm}\vrule\hspace{0.3cm}}
p{3.5cm} r
@{}}
\toprule
\textbf{Category} & \textbf{Count} &
\textbf{Category} & \textbf{Count} &
\textbf{Category} & \textbf{Count} &
\textbf{Category} & \textbf{Count} \\
\midrule

\textbf{Gender} &  &
\textbf{\mbox{Experience (years)}} &  &
\textbf{Salary per month (USD)} &  &
\textbf{Data work type} &  \\
\hdashline
Female & 57 & <1   & 5  & <50      & 15 & Labelling data                   & 120 \\
Male   & 77 & 1--3 & 44 & 50--100  & 12 & Data cleaning                    & 52 \\
       &    & 4--6 & 70 & 100--200 & 35 & Content moderation               & 67 \\
       &    & 7--10& 15 & 200--300 & 42 & Verification of user preferences & 33 \\
       &    &      &    & 300--400 & 17 &                                  &     \\
       &    &      &    & 400--500 & 12 &                                  &     \\
       &    &      &    & >500     & 1  &                                  &     \\
\midrule

\textbf{Age} &  &
\textbf{\mbox{Hours per week}} &  &
\textbf{\mbox{Exposure to graphic content}} &  &
\textbf{Big tech companies} &  \\
\hdashline
<20    & 1  & <10    & 7  & Never           & 45 & Google        & 24 \\
20--30 & 87 & 10--20 & 13 & Sometimes       & 4  & Microsoft     & 13 \\
30--40 & 45 & 20--30 & 8  & Every other day & 25 & Facebook/Meta & 35 \\
>40    & 1  & 30--40 & 36 & Every day       & 60 & OpenAI        & 51 \\
       &    & 40--60 & 65 &                 &    & Amazon        & 10 \\
       &    & >60    & 5  &                 &    &               &    \\
\bottomrule
\end{tabular}
\end{table}
}

\section{Additional Analyses}

\subsection{Household Electricity Equivalents compared to Data Centres}
\label{app:wisconsin}

The projected household electricity equivalent for the Microsoft Fairwater data centre in Wisconsin with 3.33 GW capacity is approximately 3.3 million homes. This figure is derived by calculating the annual energy output of the facility and dividing it by the average yearly electricity consumption of a Wisconsin residence. A 3.33 GW facility operating at a standard 90\% load factor which is typical for high-density hyper-scale AI infrastructure, generates roughly $3.33\times 365 \times 24 \times 0.9 = 26.25$  billion kWh of electricity annually. 

Given that the average Wisconsin household consumes approximately 660 kWh per month~\cite{EIA2025}, or roughly 7,920 kWh per year, the total electrical demand of this single data centre is equivalent to the consumption of approximately 3,315,000 households.

Studies also show that data centres in the US emitted over 105 million tCO$_2$eq in 2024, with carbon intensity $\approx$48\% higher than the national average, as many data centres are situated in areas with fossil-heavy grids~\citep{guidi_USdatacenters_2024}. This also mirrors the global trend, as the concentration of data centres is higher in fossil-heavy grids such as those in the US and China (see Figure~\ref{fig:datacenters_worldmap}).

\subsection{Model for Storage-related Emissions}
\label{app:storage-emissions}

 To approximate storage-related emissions, we use a simple lower-bound calculation. Provider-side energy use for dataset $i$ is estimated as
\begin{equation}
    E_{\text{prov},i} = \epsilon \times T_i \times S_{\text{prov},i},
\end{equation}
where $\epsilon = 60~\mathrm{kWh~TB^{-1}~year^{-1}}$ is the assumed storage energy intensity~\citep{raghav2025sustainable-ai}, $T_i$ is the time from creation to 1 December 2025 (years), and $S_{\text{prov},i}$ is the used storage size (TB) assumed to be constant. This provides a conservative estimate, as it omits internal replication by cloud providers. 

We apply a parallel calculation for user-side storage by assuming that a fraction of downloads result in local copies that persist for some period. We estimate user-side energy for dataset $i$ as
\begin{equation}
E_{\text{user},i} = \epsilon \times f_{\text{stored}} \times T_{\text{user}} \times D_i \times S_{\text{user},i}, 
\end{equation}
with 
$f_{\text{stored}}=0.1$ denoting the assumed share of downloads that remain stored, $T_{\text{user}}=0.25$ years the assumed average retention time, $D_i$ the total number of downloads, which we treat as uniformly distributed over the dataset’s lifetime, and $S_{\text{user},i}$ is the local dataset size. These assumptions reflect limited visibility into actual user-side behaviour and should be interpreted as indicative rather than exact. Based on the available attributes in Table~\ref{tab:HF_metadata}, 91\% of datasets are included for the provider-side estimate and 75\% datasets included for the user-side estimate.
\looseness=-1

For provider-side storage, emissions are derived using the average US grid intensity of 384 gCO$_2$eq/kWh~\citep{owid_carbon_intensity_2023}, as 552,713 of the datasets with region information (99.8\%) are hosted in the US, compared to only 316 in the EU. This corresponds to approximately 23.0 kgCO$_2$eq/TB. 
For user-side storage, we apply the global average grid intensity of 473 gCO$_2$eq/kWh~\citep{owid_carbon_intensity_2023}, given the lack of geographic information on dataset downloads, yielding an estimate of roughly 28.4 kgCO$_2$eq/TB.

\subsection{Internet Traffic and Data Curation}
\label{app:broadband}

To contextualise global data flows and digital presence, we draw on regionally disaggregated mobile and fixed broadband traffic statistics from the International Telecommunication Union (ITU)~\citep{itu2024measuring}. We use this data to illustrate structural asymmetries in global data generation and circulation that underpin large-scale AI systems. 

Digital participation is uneven well before data is curated for AI. 
Although mobile broadband networks now cover 96\% of the global population, around one third of people still do not use the internet~\citep{itu2024measuring}. This disparity is reflected in global traffic patterns. 

\begin{figure}[h!]
    \centering
    \includegraphics[width=0.7\linewidth]{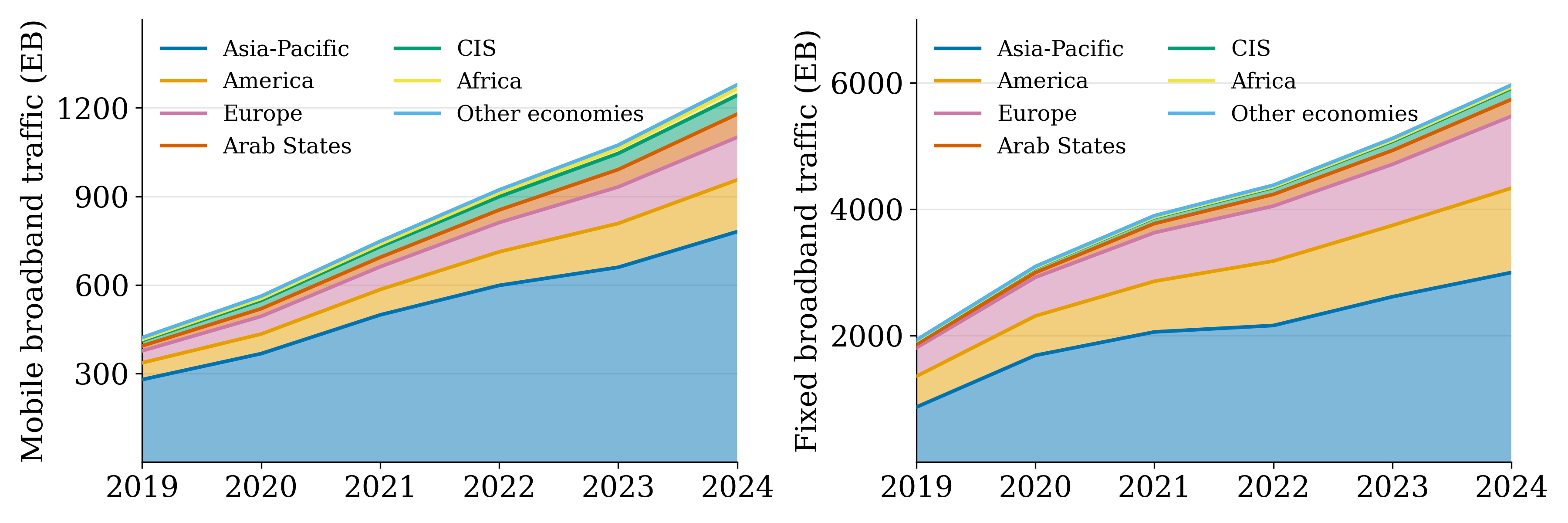}
    \caption{Mobile and fixed broadband traffic for the Asia-Pacific region, America, Europe, the Arab States, and the Commonwealth of Independent States (CIS) from 2019 to 2024. 
    Data from \citet{itu2024measuring}.  
    }
    \label{fig:broadband_traffic_by_region}
    \Description{Regional distribution of global internet traffic over time. The figure shows steady growth in both mobile and fixed broadband traffic across most regions from 2019 to 2024. Traffic volumes are dominated by the Asia-Pacific region, followed by the Americas and Europe, while Africa, the Arab States, and the Commonwealth of Independent States contribute substantially smaller shares throughout the period.}
    \vspace{-0.5cm}
\end{figure}

As Figure~\ref{fig:broadband_traffic_by_region} shows, Asia-Pacific, America, and Europe dominate both mobile and fixed broadband traffic, while Africa, the Arab States, and the Commonwealth of Independent States (CIS) generate comparatively little. 

Even though, traffic volume is not a direct measure of AI data contribution, it provides a meaningful indication of whose digital activity is more likely to be captured, indexed, or incorporated into large-scale datasets.

\end{document}